\documentclass{emulateapj}

\begin{document}

\title{A possible physical connection
between helium-rich stellar
populations of massive globular clusters and 
the UV upturn of galactic spheroids}

\author{Kenji Bekki} 
\affil{
ICRAR,
M468,
The University of Western Australia
35 Stirling Highway, Crawley
Western Australia, 6009, Australia
}

\begin{abstract}

We discuss a possible physical connection
between   helium-rich ($Y \ge 0.35$)
stellar populations of massive globular clusters (GCs) and 
the ultraviolet (UV) upturn of galactic spheroids by using
analytical and numerical models.  In our model,
all stars are initially formed as bound or unbound star clusters (SCs)
formed from giant molecular clouds (GMCs)
and the  SCs can finally  become GCs, open clusters, and field stars
depending on physical properties of their host GMCs.
An essential ingredient of the model is that
helium-rich stars are formed almost purely
from gas ejected from  massive asymptotic giant branch
(AGB) stars.
The helium-rich star formation is assumed to occur within massive SCs
if the masses of the progenitor  GMCs are larger than
a threshold mass ($M_{\rm thres}$).
These  massive SCs
can finally become  either massive GCs 
or helium-rich field stars 
depending on whether they are disintegrated or not.
Using this model,
we show that if the initial mass functions (IMFs) in galactic spheroids
are mildly top-heavy,
then the mass fractions  of helium-rich main-sequence stars ($F_{\rm He}$)
can be as large as $\sim 0.1$
for $M_{\rm thres}=10^7 {\rm M}_{\odot}$.
$F_{\rm He}$ is found to depend on IMFs and $M_{\rm thres}$ 
such that it can be larger for shallower IMFs and smaller $M_{\rm thres}$.
The inner regions of galactic spheroids show larger $F_{\rm He}$ 
in almost all  models.
Based on these results, we suggest that
if the UV upturn of elliptical galaxies is due to the larger fractions
of helium-rich stars,
then the origin can be closely associated with top-heavy IMFs in the galaxies.
\end{abstract}


\section{Introduction}

Recent observational and theoretical studies of the Galactic GCs have suggested that
some of the massive GCs ($\omega$ Cen and NGC 2808)
have significant fractions of helium-rich stars
(e.g., Bedin et al. 2004; Norris 2004; Lee et al. 2005a; 
Piotto et al. 2005, 2007;
Renzini 2008).
The observed unusual level of helium enhancement ($\Delta Y/\Delta Z \approx 70$;
Piotto et al. 2005) and sizable fractions of helium-rich stars
in these GCs have driven many theoretical studies
to make great efforts in understanding where and how helium-rich stars can be formed
(e.g., Bekki \& Norris 2006; D'Antona \& Ventura  2007; D'Ercole et al. 2008).
As discussed by Renzini (2008),  helium-rich stars can be formed
from gaseous ejecta from massive AGB stars with initial masses ($m_{\rm I}$) ranging
from $3{\rm M}_{\odot}$ to $8{\rm M}_{\odot}$,
though the ejecta should not be diluted by interstellar gas (ISM)
to keep the original high helium abundances of the ejecta.

Code \& Welch (1979) investigated the integrated light of seven elliptical and S0 galaxies 
for a wavelength range of  1550-4250 \AA \,
 and found that some of them show  
an increase in energy at the shortest wavelengths.
Since this discovery of the UV upturn,
the origin of the UV upturn
has been extensively discussed both in observational and theoretical studies
(e.g., Bertola et al. 1980; Burstein et al. 1988; Greggio \& Renzini 1990; 
Horch et al. 1992; Dorman et al. 1995;
Brown et al. 1997; O'Connell 1999; Yi et al. 1997, 1998, 2005).
Although there can be a number of stellar candidates responsible for
the UV upturn in galaxies,
one of the most promising ones  is old  horizontal-branch stars
(e.g., Yi 2008 for a recent review).
It is  suggested that enhanced helium abundances can play an important
role in the formation of hot stars responsible for the UV upturn (Yi 2008).

Brown et al. (2003) investigated the UV emission in eight early-type galaxies
at $z=0.33$ (a look-back time of 3.9 Gyr)
and found that the UV emission in these galaxies is significantly
weaker than it is in the current epoch.
Observational studies by {\it Galaxy Evolution Explorer (GALEX)}
investigated the UV properties  of bright cluster galaxies (BCGs) in clusters
at $z<0.2$ and compared them with those of nearby giant elliptical galaxies
(e.g., Lee et al. 2005b; Ree et al. 2007). 
Ree et al. (2007)  concluded that the observed evolution of FUV $-$ $V$ color
with a model in which the dominant FUV source is hot horizontal-branch stars
(e.g., Ree et al. 2007).  
Yi et al. (2011) investigated correlations between the strength of the UV upturn
and global galactic  properties (e.g., luminosity) in BCGs
and did not find any remarkable correlations 
(see also Loubser \& S\'anchez-Bl\'azquez 2011 for  similar results).
Yi et al. (2011) therefore concluded that 
the helium sedimentation scenario proposed by Peng \& Nagai (2009) can not be supported
by their  observational results.

These observational and theoretical studies appear to suggest that 
there can be helium-rich stellar populations in diverse objects
with dramatically different masses and sizes
ranging from GCs to BCGs. Although it would be likely
that different astrophysical objects with helium-rich stars have different origins,
it would be  possible that they can have a common origin.
Previous theoretical studies, however, have not yet provided  an unified
picture for the origins of helium-rich stars in GCs, galactic bulges,
and elliptical galaxies with the UV-upturn. It is therefore worthwhile to
construct a  model to discuss the origins of helium-rich stars observed in
GCs and galactic spheroids
in a self-consistent manner.

The purpose of this paper is to present a new  model 
which can provide a possible explanation for  
the origins of helium-rich stars with $Y \ge 0.35$
in GCs and galactic spheroids in a self-consistent
manner. Based on the model, we mainly discuss what types of IMFs
are required to explain the observed possible fraction
of helium-rich stars in galactic spheroids with
the UV upturn (e.g., Chung et al. 2011).
The most important ingredient of the  model is that helium-rich
stars are formed almost purely from gas ejected from massive AGB stars with 
$3 {\rm M}_{\odot} \le m_{\rm I} \le 8 {\rm M}_{\odot}$ 
with no/little dilution of the gas with ISM.
This formation of new stars from AGB ejecta have been included also
in recent theoretical models of GC formation (e.g., D'Ercole et al. 2008).

In the present study, we assume that such formation of helium-rich stars
almost directly from AGB ejecta can be possible only within SCs formed
from massive GMCs.
By considering that the vast majority of stars are observed to be initially formed as
bound or unbound SCs (e.g., Lada \& Lada 2003), we assume that all stars 
form as SCs from GMCs in the model. Some massive SCs can retain gaseous ejecta
from AGB stars and consequently have new helium-rich stars formed form the ejecta,
and the helium-rich stars can become field stars
of their host galaxies  when the SCs become disintegrated. 
Such massive SCs can finally become GCs or nuclear SC (or stellar galactic nuclei)
depending on their birth places, if they are not destroyed by their host galaxies.
The IMFs and GMCMFs are key parameter that can determine
mass fractions of helium-rich stars in galaxies.
By using this new model, we discuss the origin of 
physical properties of the Galactic GCs and
the UV upturn in galactic bulges
and elliptical galaxies. 

Although it remains less clear how much fractions of stars 
should be helium-rich stars in galactic spheroids with  the UV upturn (Yi 2008),
Chung et al (2011) have recently suggested that if about 11\% of
stellar populations 
in galactic spheroids
are helium-rich stars,
then they can show the UV upturn for a given IMF.
In their models, they considered that the major source of far-UV flux 
originates from metal-poor and helium-enhances hot horizontal branch stars.
We therefore consider that 
if the fractions of helium-rich stars in galactic spheroids
are  less than  $\sim 0.1$, the spheroids are unlikely to show the UV upturn
in the present models. We investigate in what physical conditions (e.g., IMFs)
the fractions of helium-rich stars can be  above 0.1 for
galactic spheroids to show clearly the  
UV upturn.

A possible physical link between helium-rich stellar populations and
the UV upturn of galactic spheroids has been already pointed out
by a number of authors (e.g., Yi 2008).
Thus the main point of the present study  is  {\it not} to propose the
importance of helium-rich stellar populations in the origin of the UV
upturn {\it but to discuss how galactic spheroids
can have significant fractions  of helium-rich stellar populations.}
It should be also
stressed that the present model is idealized and less realistic
in some points (e.g., no chemical evolution of galaxies): the model should
be regarded as a first step toward better understanding the origin of helium-rich
stars in GCs and galaxies. More sophisticated numerical simulations 
including SC formation processes in galaxies will need to be done
in our future studies to address the UV upturn problem in a much more 
quantitative way.

The plan of the paper is as follows: in the next section,
we describe a  model which enables us to estimate (i) the mass fraction
of helium-rich stars in a single GC and  
(ii) the mass fraction of helium-rich main-sequence (MS) stars in a galaxy for
a given IMF and GMCMF.
In \S 3, we
present the results on the number/mass fraction of massive GCs with helium-rich
stars in the Galaxy and the mass fractions of helium-rich MS stars in galaxies
with different IMFs and GMCMFs
In \S 4,
we provide important implications of the present results in terms of
the origin of helium-rich stars in GCs, the Galactic bulge, and elliptical galaxies. 
We also discuss the origin of
the radial gradients of helium-rich stars in galactic spheroids in this 
section. 
We summarize our  conclusions in \S 5.

\section{The model}

\subsection{A cluster disintegration scenario (CDS)}
\subsubsection{SCs as fundamental building blocks of galaxy formation}

We adopt a scenario in which (i) all stars in galaxies are formed
as bound or unbound SCs from GMCs
and (ii) helium-rich stars  with $Y\ge 0.35$
can be formed from gas ejected from
AGB stars only in  massive SCs.
Some SCs can become field stars in their host galaxies after disintegration
and others can become massive
GCs or nuclear star clusters (or stellar nuclei) in galaxies.
Furthermore, if massive SCs with helium-rich stars  are destroyed by their host galaxies,
then the stars  become helium-rich field populations  in the galaxies.
Therefore,  formation efficiencies of massive SCs and IMFs in GCs and galaxies are key
parameters which can determine  their mass fractions of helium-rich stars. 
This scenario is referred to as 
``cluster disintegration scenario'' (CDS) just for convenience in 
the present study. 
Since a number of possibly unfamiliar acronyms are used in the present 
study,  their physical meanings are briefly 
summarized in  Table 1 in order for readers
to understand more clearly the present paper.

\subsubsection{Two stellar populations HNSs and HRSs with different $Y$}

In the CDS,  
stars formed directly from GMCs  are assumed to have ``normal'' $Y$ that is
observed in HII regions of galaxies 
(e.g., Peimbert, Luridiana \& Peimbert 2007) and predicted
from canonical chemical evolution (i.e., $\Delta Y/\Delta Z \approx 2$) and they
are referred to as ``HNSs'' (helium-normal stars).  After these first generation of
stars form, then the second generation of stars can form from gaseous ejecta of
AGB stars among HNSs  without dilution of the ejecta with 
ISM. These second generation of stars can show large degrees
of helium enhancement  (i.e., $\Delta Y/\Delta Z \ge  4$)
which can not be expected in canonical chemical evolution models in which
ISM and gaseous ejecta from stars are assumed to well mix and then form new stars.
These second generation of stars are referred to as ``HRSs'' 
(helium-rich stars) just for convenience. 
As demonstrated by many recent theoretical studies (e.g., D'Antona et al. 2010),
massive AGB stars can ejecta gas with enhanced helium abundances ($\Delta Y \approx 0.07$).
Therefore  it is quite reasonable to assume that HRSs can have high helium abundances.
Figure 1 illustrates a whole picture for the formation histories
of two different populations (HNSs and HRSs) from GMCs in the CDS.

Our recent numerical simulations on secondary star formation from AGB ejecta in 
SCs have shown that the ejecta can be converted efficiently into new stars only in
massive SCs with masses ($m_{\rm sc}$) larger than $\sim 10^6 {\rm M}_{\odot}$
(Bekki 2010; 2011). This is because
AGB ejecta can be accumulated in the deep potential wells of
massive SCs to form high-density gaseous regions
with the densities exceeding the threshold gas density (${\rho}_{\rm thres}$)
for star formation
(${\rho}_{\rm thres} \ge 10^4 - 10^5$ atoms cm$^{-3}$). 
We accordingly consider that AGB ejecta can be converted into new stars without
the dilution with ISM if some physical conditions are met for SCs.
We assume that (i) $m_{\rm sc}/m_{\rm gmc}$ is constant for all SCs
and (ii)  HRS formation is possible within SCs only if
the masses ($m_{\rm gmc}$) of their host GMCs can exceeds a threshold mass
($M_{\rm thres}$). The introduction of $M_{\rm thres}$ is therefore consistent with
recent numerical simulations by D'Ercole et al. (2008) and Bekki (2011).

\subsubsection{$Y$ in HRSs}

It depends on chemical  yields of AGB stars
whether HRSs can have $Y \ge 0.37$ expected for blue main-sequence (MS)
stars in $\omega$ Cen and NGC 2808 (e.g., Renzini 2008)
and for the UV upturn in elliptical galaxies
(e.g., Yi 2008). Although D'Antona et al. (2005) suggested that more massive AGB stars
with $m_{\rm s} \ge 6 {\rm M}_{\odot}$ can have $Y=0.40$,
AGB stars with lower $m_{\rm s}$ ($\le 5 {\rm M}_{\odot}$) 
could have smaller Y (see Renzini 2008 for a more detailed discussion on this).
It should be noted here that Portinari et al. (2010) suggested helium-rich populations
in GCs to have $Y \approx 0.3$ rather than $Y\approx 0.4$.  
Considering these recent results,
we adopt a model in which HRSs formed from gaseous ejecta of AGB stars with
$3 {\rm M}_{\odot} \le m_{\rm s} \le 8 {\rm M}_{\odot}$ can become helium-rich
populations of GCs and galactic spheroids.

As discussed in our previous study (Bekki \& Norris 2006),
HRSs with higher $Y$ can be formed from ejecta of Type II supernovae, if the
ejecta does not mix with ISM. 
We however does not discuss  whether and how  HRSs can be formed from 
the ejecta of Type II supernovae
in the present study.
Some observational results suggest that
the helium-enrichment parameter $\Delta Y/\Delta Z$ can be larger than 4
(e.g., $5.3 \pm 1.4$ for local K-dwarf stars of  the Galaxy; 
Gennaro et al. 2010). Therefore, it would be
possible that even in canonical chemical evolution models, metal-rich stars with 
$Z \sim 0.03$
can have large $Y$ ($ \ge  0.35$) for the primordial helium content of $Y=0.24$. 
Since we do not include chemical evolution of GCs and galaxies in the present study,
we do not discuss the above possibility either.

\subsection{IMF}

We consider that HNSs and HRSs  are formed 
in SCs with different IMFs.
The adopted IMF in number is defined as follows:
\begin{equation}
\psi (m_{\rm I}) = m_{sc,0}{m_{\rm I}}^{-\alpha},
\end{equation}
where $m_{\rm I}$ is the initial mass of
each individual star and the slope $\alpha =2.35$
corresponds to the Salpeter IMF.
The normalization factor $m_{sc,0}$ is a function of $m_{\rm sc}$,
$m_{\rm l}$ (lower mass cut-off), and $m_{\rm u}$ (upper mass cut-off):
\begin{equation}
m_{sc,0}=\frac{m_{\rm sc}
\times (2-\alpha)}{{m_{\rm u}}^{2-\alpha}-{m_{\rm l}}^{2-\alpha}}.
\end{equation}
where $m_{\rm l}$ and $m_{\rm u}$ are  set to be free parameters
in the present study.
The IMF slopes for HNSs and HRSs are denoted as $\alpha_1$ and $\alpha_2$ respectively,
and $m_{\rm l,1}$ ($m_{\rm l,2}$) and $m_{\rm u,1}$
($m_{\rm l,2}$) are for HNSs (HRSs).

Gaseous ejecta from AGB stars with masses ranging from $m_{\rm l, agb}$
to $m_{\rm u, agb
}$ can contribute to the formation of HRSs
and $m_{\rm l, AGB}$ and  $m_{\rm u, agb}$ are fixed
at $3{\rm M}_{\odot}$ and $8{\rm M}_{\odot}$, respectively.
The total mass of AGB ejecta within a SC ($M_{\rm agb}$)
is accordingly described as:
\begin{equation}
M_{\rm AGB}=\int_{m_{\rm l,agb}}^{m_{\rm u,agb}} m_{\rm ej}
\psi (m) dm,
\end{equation}
where $m_{\rm ej}$ describes the total gas mass ejected from
an  AGB star with initial mass  $m_{\rm I}$
and final mass  $m_{\rm F}$.
We derive an analytic form of $m_{\rm ej}$
($=m_{\rm I}-m_{\rm F}$) from  the observational data
by Weidemann (2000) by using the least-square fitting method, and
find:
\begin{equation}
m_{\rm ej} =0.916 m_{\rm I}-0.444.
\end{equation}
Using these equations, we estimate the mass fraction of AGB ejecta  
($f_{\rm m, agb}$) in each individual SC for a given set of IMF parameters.
$f_{\rm m, agb}$ is defined as follows:
\begin{equation}
f_{\rm m, agb} = \frac {M_{\rm agb}}{m_{\rm sc}}.
\end{equation}

In order to discuss an increase of helium mass in a SC ($\Delta M_{\rm He}$) due to
gas ejection of massive AGB stars, we adopt a formula used by Renzini (2008):
\begin{equation}
\Delta M_{\rm He} = 0.15(m_{\rm I}-3) {\rm M}_{\odot}.
\end{equation}
We estimate the mass fraction of the fresh helium gas ($f_{\rm m, He}$) in
each individual SC as follows:
\begin{equation}
f_{\rm m, He}=\frac { \int_{m_{\rm l,agb}}^{m_{\rm u,agb}} \Delta M_{\rm He}
\psi (m) dm } { m_{\rm sc} }.
\end{equation}
It is suggested that $f_{\rm m, He}$ can be 0.7 per cent
for a Salpeter IMF and $m_{\rm l}=0.5 {\rm M}_{\odot}$ (Renzini 2008).
If  super AGB stars with $8\le m_{\rm I} \le 10 {\rm M}_{\odot}$ can also
contribute to the production of helium-rich gas, then $f_{\rm m, He}$ can be 0.009
(Renzini 2008). These possibly small $f_{\rm m, He}$ made several authors to
suggest that the original GCs are much more massive than the present ones
(e.g., Bekki \& Norris 2006).

\subsection{Mass fractions of HRSs in GCs}

We investigate the mass fractions of HRSs on the MS (simply referred
to as ``MS HRSs'') in the present GCs ($f_{\rm He}$)
In order to calculate the MS
turn-off mass ($m_{\rm TO}$),
we use the following formula
(Renzini \& Buzzoni 1986):
\begin{equation}
\ln m_{\rm TO}(t_{\rm s})
= 0.0558 (\log t_{\rm s})^2 - 1.338 \log t_{\rm s} + 7.764,
\end{equation}
where $m_{\rm TO}$  is in solar units and time $t_{\rm s}$ in years.
We assume that ages of GCs and galactic spheroids are 12 Gyr
and thus $m_{\rm TO}=0.885 {\rm M}_{\odot}$.  Although the present results
can hardly depend on age differences
of $\sim 300$ Myr (corresponding to the TO epoch of
stars with $m_{\rm I}=3{\rm M}_{\odot}$) between HNSs and HRSs, we consider that
$m_{\rm TO}=0.885 {\rm M}_{\odot}$ for HNSs and 
$m_{\rm TO}=0.889 {\rm M}_{\odot}$ for HRSs.

The initial total mass of HRSs in a SC ($m_{\rm sc,2}$)
is given as:
\begin{equation}
m_{\rm sc, 2}={\epsilon}_{\rm sf, 2} M_{\rm AGB},
\end{equation}
where ${\epsilon}_{\rm sf, 2}$ 
is the star formation efficiency for stars formed from AGB ejecta.
For a  strongly bound SC to be formed,  the SC should not lose a significant
fraction of gas left behind from star formation:
${\epsilon}_{\rm sf, 2}$ needs to be larger than
0.5 (e.g., Hills 1980). We here adopt ${\epsilon}_2=1.0$ in the present study.
The total mass of MS HNSs ($m_{\rm sc, mshns}$) in a SC is given as:
\begin{equation}
m_{\rm sc, mshns}=\int_{m_{\rm l,1}}^{m_{\rm TO}} 
m\psi (m) dm.
\end{equation}
The total mass of MS HRSs ($m_{\rm sc, mshrs}$) in a SC as follows:
\begin{equation}
m_{\rm sc, mshrs}={\epsilon}_{\rm sf, 2} \int_{m_{\rm l,2}}^{m_{\rm TO}} 
m\psi (m) dm,
\end{equation}
where the normalization factor
of the IMF ($\psi$) is determined by $m_{\rm sc, 2}$. Therefore,
the mass fraction of MS HRSs in a SC is given as:
\begin{equation}
f_{\rm He}= \frac { m_{\rm sc, mshrs} } {  m_{\rm sc, mshns} +  m_{\rm sc, mshrs} }.
\end{equation}
Thus $f_{\rm He}$ depends on $\alpha_1$, $\alpha_2$, 
$m_{\rm l,1}$, $m_{\rm u,1}$,
$m_{\rm l,2}$, and $m_{\rm u,2}$.

\subsection{GMCMF}
In order to discuss the number/mass fraction of MGCs in the GC system (GCS) of
the Galaxy and mass fractions of HRSs in galactic spheroids, 
we introduce GMCMFs.
A galactic spheroid is initially composed of numerous GMCs from which
unbound and bound SCs can be formed.
We adopt the following  GMCMF for GMCs:
\begin{equation}
\Psi (m_{\rm gmc}) = M_{g,0}{m_{\rm gmc}}^{-\beta},
\end{equation}
where  $\beta$ describes the slope of a GMCMF and 
is observed to be $1.6-1.7$ for galaxies in the Local Group (Rosolowski  2005).
The normalization factor $M_{g,0}$ is a function of the total
gas mass in a galactic spheroid ($M_{\rm g}$),
$m_{\rm gmc, l}$ (lower mass cut-off), and $m_{\rm gmc, u}$ (upper mass cut-off):
\begin{equation}
M_{g,0}=\frac{M_{\rm g}
\times (2-\beta)}{{m_{\rm gmc, u}}^{2-\beta}-{m_{\rm gmc, l}}^{2-\beta}}.
\end{equation}
where $m_{\rm gmc, l}$ and $m_{\rm gmc, u}$ are set to be
$10^3 {\rm M}_{\odot}$ and $10^8 {\rm M}_{\odot}$, respectively, 
in the present study.

We consider that $\beta$ could be different in different regions of a galaxy
and in different galaxies and thus investigate models with different $\beta$
($1.1 \le \beta \le 2.0$). We assume that star formation efficiencies 
(${\epsilon}_{\rm sf, 1}$)
within host GMCs are constant, though they can be different between different GMCs.
This is because we focus on mass fractions of HRSs in GCs and galactic spheroids and
need to more clearly show their dependences on IMFs and GMCMFs.
The present results do not depend so strongly on  $m_{\rm gmc, l}$ and
$m_{\rm gmc, u}$ for reasonable ranges of these two.

\subsection{Formation efficiency of bound SCs in GMCs}

Some SCs get disintegrated and consequently become
field stars in their host galaxies and others
survive from disintegration and finally become GCs. 
SCs with HRSs can finally become massive GCs (like $\omega$ Cen and NGC 2808), 
which are referred to
 ``MGCs'': other GCs with no HRSs are simply referred to as GCs.
The total mass of SCs with HRSs formed 
in a galaxy is a key parameter which determines the total helium mass
of the  galaxy.  Therefore,  the formation efficiency of 
{\it bound} SCs  (${\epsilon}_{\rm bsc}$) 
in GMCs is one of the most important parameter in the present study.
Here ${\epsilon}_{\rm bsc}$ is defined such that if ${\epsilon}_{\rm bsc}=1$ (0.1),
then each (one in ten) GMC 
can form a SC that is strongly bound so as to form stars from AGB ejecta.  

Although it remains observationally unclear what a reasonable value is for 
${\epsilon}_{\rm bsc}$,  our previous simulations showed that
formation efficiencies of GCs in starbursting galaxy mergers can be much higher
in comparison with isolated disk galaxies owing to rather high pressure
of ISM (Bekki et al. 2002). The simulations suggested that gas can be converted into
strongly bound GCs rather 
than field stars (or weekly bound SCs) in major galaxy mergers.
These results imply that if galactic spheroids are formed from major mergers,
they can have rather high ${\epsilon}_{\rm bsc}$. We discuss ${\epsilon}_{\rm bsc}$
later in \S 4.

\subsection{Number and mass fractions of MGCs in the Galaxy}

We  discuss number and mass fractions
of MGCs ($F_{\rm n, mgc}$ and $F_{\rm m, mgc}$, respectively)
can form ``genuine GCs'' that are observed to show O-Na anti-correlations
and thus evidence for the presence of multiple stellar populations
(e.g., Carretta et al. 2010)
in the Galactic GCS.
We consider that GMCs with $m_{\rm gmc} \ge 10^6 {\rm M}$ 
(or $m_{\rm sc} \ge 10^5 {\rm M}_{\odot}$ for ${\epsilon}=0.1$)
form ``genuine GCs''.  The threshold GMC mass for the genuine GCs
is denoted as $m_{\rm gmc, gc}$ for convenience.
We estimate $F_{\rm n, mgc}$ as follows:
\begin{equation}
F_{\rm  n, mgc}= \frac { \int_{M_{\rm thres}}^{m_{\rm gmc, u}} 
  {\epsilon}_{\rm bsc} \Psi (m) dm }
{ N_{\rm gc} }, 
\end{equation}
where $N_{\rm gc}$ is the total number of genuine  GCs and
given as:
\begin{equation}
N_{\rm  gc}= \int_{m_{\rm gmc, gc}}^{m_{\rm gmc, u}} 
  {\epsilon}_{\rm bsc} \Psi (m) dm .
\end{equation}
Likewise, 
$F_{\rm m, mgc}$ is estimated as follows:
\begin{equation}
F_{\rm  m, mgc}= \frac { \int_{M_{\rm thres}}^{m_{\rm gmc, u}} 
 {\epsilon}_{\rm sf, 1}  {\epsilon}_{\rm bsc}  m\Psi (m) dm }
{ M_{\rm gc} }, 
\end{equation}
where $M_{\rm gc}$ is the total stellar mass of the genuine GCs and given as:
\begin{equation}
M_{\rm  gc}= \int_{m_{\rm gmc, gc}}^{m_{\rm gmc, u}} 
 {\epsilon}_{\rm sf, 1}  {\epsilon}_{\rm bsc} m\Psi (m) dm .
\end{equation}

\subsection{Mass fractions of HRSs in galactic spheroids}

The total stellar mass of a galactic spheroid ($M_{\rm s}$) is estimated
as follows:
\begin{equation}
M_{\rm  s}=\int_{m_{\rm gmc, l}}^{m_{\rm gmc, u}} 
{\epsilon}_{\rm sf, 1} m\Psi (m) dm.
\end{equation}
The total mass of HRSs in the  galactic spheroid ($M_{\rm s,hrs}$) is estimated
as follows:
\begin{equation}
M_{\rm  s,hrs}=\int_{M_{\rm thres}}^{m_{\rm gmc, u}} 
{\epsilon}_{\rm sf, 1} {\epsilon}_{\rm sf, 2}
{\epsilon}_{\rm bsc}
 f_{\rm m, agb} 
m\Psi (m) dm.
\end{equation}
In these equations, the term ${\epsilon}_{\rm bsc}$ is included only in the equation
(20), 
because whether SCs are bound or unbound does not matter in estimating the
total stellar mass $M_{\rm s}$.
The mass fraction of HRSs in a galactic spheroid ($F_{\rm He, t}$) is therefore
given as follows:
\begin{equation}
F_{\rm  He, t}=\frac{ M_{\rm s,hrs} } { M_{\rm s} }.
\end{equation}
Although this $F_{\rm He, t}$ enables us to understand how much fraction of stars can be
HRSs in a galactic spheroid, it is different from $F_{\rm He}$ which is more useful
when the origin of galactic spheroids with HRSs are discussed.

The total mass of MS stars in a galactic spheroid ($M_{\rm s, ms}$) is estimated
as follows:
\begin{equation}
M_{\rm  s, ms}=\int_{m_{\rm gmc, l}}^{m_{\rm gmc, u}} 
{\epsilon}_{\rm sf, 1} f_{\rm ms, 1} m\Psi (m) dm,
\end{equation}
where $f_{\rm ms, 1}$ is the mass fraction of MS stars among all stars
and depends on the IMF parameters of HNSs (e.g., $\alpha_1$).
The total mass of MS HRSs ($M_{\rm s,mshrs}$) is estimated
as follows:
\begin{equation}
M_{\rm  s,mshrs}=\int_{M_{\rm thres}}^{m_{\rm gmc, u}} 
{\epsilon}_{\rm sf, 1} {\epsilon}_{\rm sf, 2}
{\epsilon}_{\rm bsc}
 f_{\rm ms, 2} 
m\Psi (m) dm,
\end{equation}
where $f_{\rm ms, 2}$ is the mass fraction of MS HRSs among all stars formed 
as HRSs and determined by IMF parameters of HRSs. 
Therefore  $F_{\rm He}$ is given as follows:
\begin{equation}
F_{\rm  He}=\frac{ M_{\rm s,mshrs} } { M_{\rm s, ms} }.
\end{equation}

In the present study, ${\epsilon}_{\rm sf, 1}$ 
is assumed to be independent of $m_{\rm gmc}$
and accordingly the present
results on $F_{\rm He}$ do not depend on  ${\epsilon}_{\rm sf, 1}$.
Since we focus on $F_{\rm He}$ in galactic spheroids, 
it is unimportant whether low-mass SCs are unbound (or bound) to become
fields stars (open/globular clusters) after SC disintegration.
The parameter ${\epsilon}_{\rm bsc}$ is therefore unimportant for GMCs with 
$m_{\rm gmc} \le M_{\rm thres}$. We assume that
${\epsilon}_{\rm bsc}=1$ for GMCs that form MGCs in all models.
It would be possible that $F_{\rm He}$ is significantly overestimated 
owing to the adopted assumption of ${\epsilon}_{\rm bsc}=1$.
We later discuss this point in \S 4. Thus free parameters are
$\alpha_1$,
$\alpha_2$,
$m_{\rm l,1}$,
$m_{\rm l,2}$,
$m_{\rm u,1}$,
$m_{\rm u,2}$,
$\beta$,
and $M_{\rm thres}$ in the present study.
Table 2 and 3 briefly summarizes the physical meanings of these parameters
and the definition of physical quantities investigated in the present study
(e.g., $F_{\rm He}$), respectively, 
so that readers
can understand more clearly the present results.

\section{Results}

\subsection{GCs}

Figure 2 shows how $f_{\rm m, He}$ and $f_{\rm m, agb}$ depend on IMF parameters,
$\alpha_1$, $m_{\rm l,1}$, and $m_{\rm u, 1}$.  Clearly, $f_{\rm m, He}$ is
rather small ($\sim 0.005$) for a canonical IMF with $\alpha_1=2.35$,
$m_{\rm l,1}=0.1 {\rm M}_{\odot}$, 
and $m_{\rm l,u}=100 {\rm M}_{\odot}$. 
This result means  that only a small fraction of original stellar mass in a SC
can be fresh helium gas that can be used for the formation of HRSs for a
canonical IMF. 
However, $f_{\rm m, He}$ can be larger than 0.01 for top-heavy IMFs ($\alpha_1 <2$)
if $m_{\rm u}$ is smaller 
(e.g., 0.02 for $\alpha_1$ and $m_{\rm u,1}=8 {\rm M}_{\odot}$).
In extreme situations with 
$m_{\rm u, 1}=8 {\rm M}_{\odot}$ (i.e., no Type II supernova),
$f_{\rm m, He}$ can increase as the IMF slope decreases  (i.e., shallower 
or more top-heavy IMF).  Although top-heavy IMFs with 
$m_{\rm u, 1}=8 {\rm M}_{\odot}$ would be highly unlikely for most GCs,
these peculiar IMFs could be associated with the origins
of some GCs with large fractions of
HRSs. It is clear that $f_{\rm m, He}$ depends on
$m_{\rm l,1}$ such that $f_{\rm m, He}$ is larger for
larger $m_{\rm l,1}$.

The parameter dependences of $f_{\rm m, agb}$ on IMF parameters
are essentially the same with those of  $f_{\rm m, He}$.
SCs with canonical IMFs with $\alpha_1=2.35$,
$m_{\rm l,1}=0.1 {\rm m}_{\odot}$, and $m_{\rm u,1}=100 {\rm M}_{\odot}$
can have  $f_{\rm m, agb} \sim 0.08$. 
However SCs with moderately top-heavy IMFs
($\alpha \sim 2$) can show 
$f_{\rm m, agb}$ as large as 0.2, 
if $m_{\rm l,1}=0.5 {\rm m}_{\odot}$ and $m_{\rm u,1}$
are smaller than $30 {\rm M}_{\odot}$
(e.g., $f_{\rm m, agb} \sim 0.29$  for $m_{\rm l,1}=0.5 {\rm M}_{\odot}$
and  for $m_{\rm u,1}=8 {\rm M}_{\odot}$).
Given that these SCs can not
lose significant fractions of their masses owing to the IMFs with rather small
$m_{\rm u,1}$,  they  are highly likely to survive after
supernova explosions so that secondary star formation can be possible within 
the SCs.  If most of the AGB ejecta can be converted into HRSs within SCs, then
the SCs can have significant fractions of HRSs. 
Thus the combination of moderately top-heavy IMFs and rather small $m_{\rm u, 1}$
can significantly increase $f_{\rm m, agb}$, 
which suggests that SCs can contain significant fractions of HRSs without
disintegration.
The importance of such combination of $\alpha_1 \sim 2$ and low  $m_{\rm u, 1}$
in the formation of HRSs was not discussed in previous studies 
(e.g. Bekki \& Norris 2006; Renzini 2008).

Figure 3 shows how $f_{\rm He}$ depends on $\alpha_1$ for a given $\alpha_2$ in
SCs with $m_{\rm u,1}=0.1{\rm M}_{\odot}$,
$m_{\rm u,1}=100{\rm M}_{\odot}$,
and ${\epsilon}_{\rm sf,2}=1$.
Clearly, SCs with canonical IMFs with $\alpha_1=2.35$ and $\alpha_2=2.35$
have a small $f_{\rm He}$ ($\sim  0.08$). SCs with more top-heavy IMFs in their HNSs 
(i.e., smaller $\alpha_1$) show larger $f_{\rm He}$ for a given $\alpha_2$.
SCs with larger $\alpha_2$ (more ``bottom-heavy'') show larger $f_{\rm He}$
for a given $\alpha_1$. For SCs with canonical IMFs in their HRSs
($\alpha_2=2.35$)  to have
significant fractions of HRSs ($f_{\rm He} \sim 0.2$), 
they need to have top-heavy IMFs in HNSs ($\alpha_1 <2$).  
These results do not depend so strongly on $m_{\rm l, 2}$ and $m_{\rm u, 2}$,
though $f_{\rm He}$ can be larger for smaller $m_{\rm u, 2}$.
If SCs have too top-heavy IMFs (e.g., $\alpha_1 < 1.5$), then
they can lose most of their mass (more than 50\%) 
through Type II supernova explosions.
These SCs are likely to become disintegrated before secondary star formation
from AGB ejecta can proceed. Thus it is unlikely that SCs with large $f_{\rm He}$
($>0.5$) can be now observed as bound GCs.

SCs with moderately top-heavy IMFs ($\alpha_1 \sim 2$)
can have significant fractions of HRSs ($f_{\rm He} \sim 0.2$),
if these SCs can not be disintegrated due to
mass loss through  Type II supernova explosions.
It should be however stressed that
${\epsilon}_{\rm sf, 2}=1$ and no mixing of AGB ejecta with ISM are assumed
in these investigation.
Furthermore, it is assumed that all of gaseous ejecta with AGB stars with
different masses and lifetimes can be converted into new stars simultaneously.
In real environments of GC formation, 
${\epsilon}_{\rm sf, 2}$ can be significantly smaller than 1
and AGB ejecta could be diluted by ISM.
Also, massive stars among  HRSs  can explode as
supernovae and consequently prevent gaseous ejecta from AGB stars
with lower masses from being converted into new stars.
Therefore $f_{\rm He}$ would be likely to be
lower than those estimated in the present study.
It should be also  stressed that if rather small $m_{\rm u, 1}$ 
($<30 {\rm M}_{\odot}$) is adopted,  then $f_{\rm He}$ can be 
higher than those shown in Figure 3.

\subsection{The fraction of MGCs in the Galactic halo}

By assuming that all of the Galactic GCs were formed from GMCs,
we here briefly discuss how much fraction of GCs were originally formed
as MGCs with HRSs (like $\omega$ Cen and NGC 2808). In the present study,
SCs formed from GMCs with $m_{\rm gmc} \ge M_{\rm thres}$ can 
efficiently convert their AGB ejecta into new stars without dilution by
ISM so that they can have HRSs. 
Other less massive GC that evolve from SCs 
formed from GMCs with $m_{\rm gmc, gc} \le m_{\rm gmc} < M_{\rm thres}$
can finally become genuine GCs with no/few HRSs. 
SCs formed from GMCs with $m_{\rm gmc} \le m_{\rm gmc, gc}$ can not become 
genuine GCs in the present study. 
Given that the Galactic GCs with significant fractions of helium-rich
populations (e.g., $\omega$ Cen) have total masses larger than $10^6 {\rm M}_{\odot}$,
we consider that $M_{\rm thres}=10^7 {\rm M}_{\odot}$ (for ${\epsilon}_{\rm sf, 1}=0.1$)
is reasonable.

Figure 4 clearly shows that although the number fraction of MGCs is small
($F_{\rm n, mgc} \sim 0.16$) for a reasonable $\beta$ (=1.7)
and $M_{\rm thres}=10^7 {\rm M}_{\odot}$, the mass fraction ($F_{\rm m, mgc}$)
is quite significant ($>0.6$). 
Both $F_{\rm n, mgc}$ and $F_{\rm m, mgc}$ are
larger for smaller $\beta$ and smaller $M_{\rm thres}$
owing to the larger fractions of GMCs with $m_{\rm mgc} \ge M_{\rm thres}$. 
The essential reason for large $F_{\rm m, mgc}$ ($>0.5$) in some models is that GMCMFs have
slopes larger 
than $-2$ (i.e., $\beta < 2$).  
It should be stressed that ${\epsilon}_{\rm bsc}$ is assumed to 
be constant for all SCs in deriving
these results.
If host GMCs for MGCs have higher ${\epsilon}_{\rm bsc}$ than those for other GCs,
then $F_{\rm n, mgc}$ and $F_{\rm m, mgc}$ can be even larger than those shown
in Figure 4.

\subsection{Galactic spheroids}

Figure 5 shows that the mass fraction of HRSs ($F_{\rm He, t}$) in the model with
$\beta=1.7$ 
and $M_{\rm thres}=10^7 {\rm M}_{\odot}$ is rather small ($\sim 0.05$)
for canonical IMFs ($\alpha_1=2.35$).
Since ${\epsilon}_{\rm bsc}$ is assumed to be 1 for all GMCs 
with $m_{\rm gmc} \ge M_{\rm thres}$ in the present study,
$F_{\rm He, t}$ can be significantly  smaller than 0.05 in real galaxies
where ${\epsilon}_{\rm bsc}$ can be different in different
star-forming regions and  well less than 1 in some star-forming regions.
These results
mean that galactic spheroids are unlikely to have significant fractions
of HRSs for canonical IMFs ($\alpha_1=2.35$).
Clearly, $F_{\rm He, t}$ is larger for smaller $\beta$,
because there are larger fractions of  GMCs that have $m_{\rm gmc} \ge M_{\rm thres}$
and thus can host MGCs with HRSs  in galactic spheroids.
Also $F_{\rm He, t}$ is larger 
for smaller $M_{\rm thres}$ for a given $\beta$ owing to larger fractions of GMCs with
$m_{\rm gmc} \ge M_{\rm thres}$.

The mass fractions of MS HRSs ($F_{\rm He}$) can be 
significantly larger than $F_{\rm He, t}$ in
galactic spheroids, if the mass fractions of MS HNSs are smaller owing to top-heavy
IMFs. Figure 5 shows how $F_{\rm He}$ depends on $\alpha_1$ for a given set of
$\alpha_2$, $\beta$, and $M_{\rm thres}$.
As expected from Figure 4,
  $F_{\rm He}$ is rather small ($\sim 0.05$) for canonical
IMFs with $\alpha_1=2.35$ and $\alpha_2=2.35$. Irrespective of $\alpha_2$,
$\beta$, and $M_{\rm thres}$,  IMFs of HNSs in galactic spheroids
should be top-heavy ($\alpha_1 \le  2$)
for the spheroids to have significant fractions of MS HRSs ($F_{\rm He} \ge0.1$).
It should be stressed, however, that the observationally suggested
$F_{\rm He}$ ($\sim 0.1$) for galactic spheroids with the UV upturn 
(Chung et al. 2011) can be explained by the models with mildly top-heavy
IMFs of $\alpha_1 \sim 2$ that are not so exotic.

It is clear that $F_{\rm He}$ is larger for  larger $\alpha_2$ 
for a given set of $\alpha_1$, $\beta$, and $M_{\rm thres}$. This is because
larger numbers of low-mass HRSs ($m_{\rm I} < 3{\rm M}_{\odot}$) can be formed
for larger $\alpha_2$. 
Also, $F_{\rm He}$ can be larger for smaller $\beta$ and smaller $M_{\rm thres}$,
because larger numbers of GMCs with $m_{\rm mgc} \ge M_{\rm thres}$ can be formed.
These results mean  that whether galactic spheroids can have 
higher $F_{\rm He}$ ($\ge 0.1$) depends on physical properties of GMCs
and physical  processes of secondary star formation in MGCs.
It should be stressed here that both ${\epsilon}_{\rm sf,2}=1$ 
and ${\epsilon}_{\rm bsc}=1$
are assumed in these estimation: top-heavy IMFs are required for galactic spheroids
to have significant fractions of MS HRSs even in these maximum formation efficiencies
of HRSs and MGCs.

\section{Discussion}

\subsection{The lost GCs with helium-rich stars}

The Galactic GCs can sink into the inner region of the bulge owing to dynamical
friction against the halo within $\sim 13$ Gyr,
if their initial masses ($m_{\rm gc}$) is enough large  (Binney \& Tremaine 1987).
By assuming that the Galaxy has a singular isothermal sphere,
the time scale of dynamical friction for a GC ($t_{\rm df}$)  can be estimated
as follows:
\begin{equation}
t_{\rm df}= 2.3 
{ ( \frac{ \ln \Lambda} {10} )  }^{-1}
{ ( \frac{ r_{\rm i} } {2 {\rm kpc} } )  }^{2}
{ ( \frac{ v_{\rm c} } {220 {\rm pc} } )  }
{ ( \frac{ m_{\rm gc} } {10^7  {\rm M}_{\odot} }  )  }^{-1}
{\rm Gyr},
\end{equation}
where  $\lg \Lambda$, $r_{\rm i}$, $v_{\rm c}$ are
the Coulomb logarithm,  the initial distance of the GC from the Galactic center,
and the circular velocity of the Galaxy.
In the above estimation, the reference value of $m_{\rm gc}$ was set to be
a higher, because GCs can have significantly higher masses at their birth.
The above equation means that initially massive GCs with HRSs 
have already  sunk into the central region of the Galactic bulge owing to
dynamical friction, if their $r_{\rm i}$ are less than 2 kpc.
These GCs can not be observed as the Galactic halo GCs and contribute to the formation
of helium-rich stars in the bulge.

In order to estimate how much fraction of massive GCs in the GCS of the 
Galaxy has been already lost in the bulge, we consider
that the GCS has a spherical distribution with
a density profile of ${\rho}(r)$ $\propto$ $r^{-3.5}$. This is
consistent with that observed for the Galactic GCS
(Djorgovski \&  Meylan 1994). The GCS is distributed within 35 kpc of the Galaxy and  
have a half-number radius of 5 kpc. We here investigate a threshold radius $R_{\rm thres}$  
for which $t_{\rm df}$ can be smaller than 13 Gyr 
for a given $m_{\rm gc}$.  GCs with $r_{\rm i} \le R_{\rm thres}$ can spiral into
the center of bulge within 13 Gyr so that they can not be observed as the Galactic halo GCs.
It is found that $\sim 49$\% of GCs with $m_{\rm gc}=10^7 {\rm M}_{\odot}$ have been
already lost for the above adopted GCS radial profile
(i.e., $R_{\rm thres}=4.7$ kpc). The fraction of these lost GCs
($F_{\rm lost}$) is smaller for smaller $m_{\rm gc}$: $F_{\rm lost}$ is 0.37
for $m_{\rm gc}=5 \times 10^6 {\rm M}_{\odot}$ and 0.06 for
$m_{\rm gc}=10^6 {\rm M}_{\odot}$.

These results imply that what we can now observe as MGCs (e.g., $\omega$ Cen and NGC 2808)
could be [51-63]\% of original MGCs ($m_{\rm gc} \ge 5 \times 10^6 {\rm M}_{\odot}$)
formed in the Galactic halo. The lost MGCs would have
been destroyed by the Galactic bulge and consequently their HRSs  would have
been dispersed into the inner bulge region: the stars could be now observed as HRSs 
in the bulge. This selective loss of MGCs due to dynamical friction
might  well occur in galactic bulges
and giant elliptical galaxies. Sohn et al. (2006) revealed a large number of GCs
with strong UV light in M87 (see also Kaviraj et al. 2007).
If these GCs contain large fractions of HRSs,
then M87 could have already swallowed a fraction of
these possible MGCs to add their HRSs 
to its main stellar spheroid.

\subsection{Top-heavy IMFs for the origin of the UV upturn in elliptical galaxies and BCGs}

The present study has demonstrated that if IMFs are top-heavy 
(i.e., $\alpha_1 < 2$), then significant fractions ($\sim 0.1$) of
MS stars in galactic spheroids  can be 
HRSs. This is mainly because a larger amount of AGB ejecta with high helium
abundances  can be converted
into new stars without dilution of the ejecta with ISM. 
This therefore implies that 
if the UV upturn is due largely to large fractions of 
MS HRSs within them,
as observationally suggested (Chung et al. 2011),
then  galactic spheroids with mildly top-heavy IMFs are more likely 
to show the UV upturn. Then are the proposed top-heavy IMFs  consistent
with other observational properties of galactic spheroids such
as their  luminosity evolution and chemical abundances  ?

A recent observation suggested that IMFs in massive galaxies
at $0.02 \le z \le 0.83$ are significantly flatter 
(corresponding to $\alpha=1.3$ in our model thus
even shallower than the required slope of $ \sim -2$) 
than the present-day value of the Galaxy
and proposed a ``bottom-light'' IMF for massive galaxies at higher redshifts
(van Dokkum 2008).
This observational result is consistent with the present proposal
and accordingly implies  that if
the observed massive galaxies in van Dokkum (2008)
are progenitors for the present galactic spheroids,
then some of them  can show the UV upturn owing to larger $F_{\rm He}$.
It should be stressed here that the observed fraction of spheroid populations
with the UV upturn is small (e.g., Yi et al. 2011)
and thus the observed {\it mean} properties of high redshift spheroids
could be different from those of the present ones with the UV-upturn.

The required top-heavy IMFs are also consistent with the IMFs proposed 
for explaining the chemical properties of galactic spheroids 
(e.g., bulges), such as
metallicity distribution functions (MDFs) and  [$\alpha$/Fe]
(e.g., Matteucci \& Brocato 1990; Nagashima et al. 2005; Ballero et al. 2007;
Tsujimoto et al. 2010). Loewenstein (2006) showed that rapid star formation
with top-heavy IMFs in massive galaxies
are necessary to explain the observed Fe abundances
of intra-cluster medium (ICM) which contains heavy  metals from supernovae 
of massive early-type galaxies in cluster of galaxies.
The top-heavy IMFs adopted in  these previous studies suggest that
the required IMF for explaining the UV-upturn observed in some spheroids
is neither  unreasonable nor  unrealistic.
Larson (1998) suggested that gas clouds with higher temperature in
galaxies at high redshifts can be 
responsible for star formation with top-heavy IMFs. Thus the required
top-heavy IMFs could be closely associated with high-redshift formation
of the vast majority of stars in galactic spheroids.

It should be also stressed that if IMFs are too top-heavy (e.g., $\alpha_1<1.5$), then
the SCs could become disintegrated before AGB ejecta can be recycled and converted
into new stars. About 45\% in mass can become stars with masses larger 
than $8{\rm M}_{\odot}$ (thus supernova) in SCs 
for $\alpha_1=1.95$, 
$m_{\rm l,1}=0.1{\rm M}_{\odot}$, and
$m_{\rm u,1}=100{\rm M}_{\odot}$. These SCs can not become disintegrated and thus
can continue secondary star formation,
because more than 50\% of their masses can still remain in SCs
(e.g., Hills 1980).
However,
if $\alpha_1=1.5$, 
$m_{\rm l,1}=0.1{\rm M}_{\odot}$, and
$m_{\rm u,1}=100{\rm M}_{\odot}$ 
then the mass fractions of supernovae in SCs are 0.74
so that SCs can lose most of their original masses.
These SCs are highly likely to
get disintegrated shortly after supernova explosions owing to their mass loss.
Therefore, too top-heavy IMFs ($\alpha_1<1.5$) are not ideal for galactic spheroids to
have helium-rich stellar populations. 

Peng \& Nagai (2009) proposed that sedimentation of helium in clusters
of galaxies can be responsible for the formation of HRSs in BCGs.
They also predicted that the UV flux strength is stronger in more massive, low-redshift,
and dynamically relaxed BCGs.  Their predictions, however, have been recently
challenged by an observational study by Yi et al. (2011) which 
have found no correlation between the UV strength and rank/luminosities of BCGs
and showed no clear difference in UV upturn fraction or strength in BCGs.
The present study provides an alternative explanation for the origin of the 
BCGs with the UV upturn: BCGs show the UV upturn because IMFs at their formation
were top-heavy ($\alpha_1<2$) and therefore they have larger $F_{\rm He}$.
In the CDS with top-heavy IMFs,  some BCGs can not show the UV upturn,
because most stars are formed as SCs with canonical IMFs (whereas some can
owing to top-heavy IMFs).

However, the origin of UV upturn in BCGs
would not be so simple as the CDS with top-heavy IMFs  envisages.
Recent theoretical models  have shown that
hierarchical merging of smaller galaxies can play a vital role in
the formation of BCGs (e.g., De Lucia \& Blaizot 2007). The IMFs are 
observationally suggested to be  different
between faint and luminous galaxies  (e.g., Hoversten \& Glazebrook 2008).
Therefore, if BCGs were formed by numerous mergers between galaxies
with different luminosities  in clusters,
then not just IMFs but also merging histories (e.g., fractions of
faint/luminous galaxies) can be key determinants for whether BCGs can show 
the UV upturn. 

\subsection{Halo-spheroid connection }

Recently Martell et al. (2011) have revealed that about 3\% of the Galactic
halo stars can have depleted carbon and enhanced nitrogen abundances that
are very similar to the chemical abundances observed 
for the ``second-generation'' of stars in GCs. 
Since these stars with characteristic chemical abundances can be formed
from gaseous AGB  ejecta of the ``first generation of stars'' 
(e.g., Bekki et al. 2007; D'Ercole et al. 2008),
they suggested
that (i) these stars originate from GCs and
(ii)  a minimum 17\% of the present-day
mass of the Galactic stellar halo was originally formed in GCs. 
Although it is difficult 
for observational studies to directly estimate helium abundances of stars
in GCs,
Bragaglia et al. (2010) have recently estimated the average
enhancement in the helium mass fraction Y between the first and second
generation (corresponding to HNSs and HRSs, respectively, in the
present study)
for 19 GCs with the Na-O anticorrelation. The estimated
average enhancement is about $0.05-0.11$, which means that a significant
fraction of star in these GCs can have HRSs. If $\sim 10$\% of all GC
stars are HRSs, then about 1.7\% of the halo stars can be regarded
as HRSs.

This smaller number of 1.7\% implies that the Galactic halo can not be 
identified as a galactic component with the UV upturn.  
Then what mechanism is responsible for the possible large difference
in the mass fraction of HRSs ($F_{\rm He}$) between the Galactic stellar
halo and galactic spheroids ? The present CDS suggests that most stars
in the Galactic halo can originate from disintegration of low-mass
SCs with canonical IMFs whereas  those of galactic spheroids
originate from stars initially in more massive SCs with top-heavy
IMFs. Furthermore, as discussed in \S 4.1, the more massive SCs (or GCs)
with HRSs can rapidly sink into the central regions of their hosts so that
they can not contribute to  the formation of HRSs in their halo regions:
these more massive GCs can preferentially become the building blocks
of the central spheroidal components in galaxies.
Thus it is highly likely that $F_{\rm He}$ can be significantly 
different between halos and spheroids in galaxies.

\subsection{High formation efficiencies of bound massive SCs at
the epoch of spheroidal formation}

It should be stressed here that ${\epsilon}_{\rm  bsc} \sim 1$ 
has been so far assumed:
GMCs with $m_{\rm gmc} \ge M_{\rm thres}$ need
to almost always form bound SCs  that can finally form  helium-rich stars. 
If ${\epsilon}_{\ bsc} \sim 0.1$, then it is very hard for galactic spheroids
to have $F_{\rm He} \sim 0.1$ even for  very top-heavy IMFs (e.g., $\alpha_1=1$).
Only spheroids with 
$\alpha_1 \sim 1$ and $m_{\rm u,1}\sim 8 {\rm M}_{\odot}$  (i.e.,
peculiar top-heavy IMFs)
can have $F_{\rm He}$ as large as 0.1 for ${\epsilon}_{\ bsc} \sim 0.1$.
Photometric studies of super star clusters (SSCs) in a starbursting luminous infrared galaxy
Arp 220 showed that the nuclear SSCs in Arp 220 contribute to $\sim 20$\% of the total
bolometric luminosity of Arc 220 (Shioya et al. 2001).  
Larsen \& Richtler (2000) found that the formation efficiency of
SCs can be higher in local regions with high star formation rate per unit
area for disk galaxies. 
These observations imply that (i) a significant fraction of new stars
are formed as SCs in a starburst or in high-density star-forming regions
and thus (ii) galactic spheroids can have  high cluster formation 
efficiencies if they were formed from massive
starbursts in high-density environments. 

However,
owing  to the lack of extensive observational studies on formation
efficiencies of massive SCs in galaxies, 
it is currently very hard to discuss 
what a reasonable value of
${\epsilon}_{\ bsc}$ is for massive GMCs that can be progenitors
for bound SCs with HRSs.
If ${\epsilon}_{\ bsc}$ can be significantly larger than 0.1,
then the present CDS is promising as the origin of
the UV upturn in galactic spheroids, though top-heavy IMFs are  still
required. Thus, ultimately speaking, the origin of the 
UV upturn is closely related to the formation processes of MGCs
from massive GMCs at the epoch of galaxy formation in the present
CDS.

\subsection{Helium-rich populations in the Galactic bulge ?}

Nataf et al. (2011) revealed  that the metal-rich stellar populations
of the Galactic bulge can have $Y \sim 0.35$ based on the observational results
on physical properties of  the red giant branch bump (RGBB).
They provided some important implications of the metal-rich HRSs in the bulge 
and suggested that galactic bulges in general can have HRSs like the bulge.
The present results imply that if the Galactic bulge had a top-heavy IMF
at its formation, then the present stellar populations 
can have a significant fraction ($\sim 0.1$) of HRSs.  Given that previous chemical evolution
models of the Galactic bulge
(e.g., Ballero et al. 2007; Tsujimoto et al. 2010) suggested a moderately top-heavy
IMF ($\alpha \sim 2.05$) for the bulge, 
it is possible that the Galactic bulge can have a significant fraction
of HRSs owing to the top-heavy IMF.

As suggested by Nataf et al. (2011), the dominant populations of the  RGBB 
 (with [Fe/H]$>0$) in the bulge can have a rather high $Y$ ($\sim 0.35$): 
not just an $\sim 10$\% of the bulge population 
needs to have $Y \sim 0.35$ to explain their observational results.
If more than  30\% of the metal-rich stellar populations have $Y \sim 0.35$,
then the IMF slope ($\alpha$) for the populations should be well less than 1.5
for $\beta = 1.7$ and $M_{\rm thres}=10^7 {\rm M}_{\odot}$.
The required IMF is significantly more top-heavy than those suggested by
chemical evolution studies (e.g., $\alpha \sim 2.05$; Tsujimoto et al. 2010)
and thus would not explain the observed  global properties of the bulge 
such as the metallicity distribution function of the bulge stars.
However, if metal-rich stellar populations of the bulge has a very top-heavy IMF
and if other populations have moderately top-heavy IMFs,
then it would be possible that both the observed high fraction of metal-rich HRSs
and global chemical properties in the bulge can be self-consistently explained.
This possibility needs to be explored in our future studies based on detailed
chemical evolution models.

\subsection{The origin of the correlation between the strength
of UV upturn  and Mg$_2$ index}

Recent observational studies of 48 nearby early-type galaxies by the 
SAURON project (Bureau et al. 2011) have confirmed the presence 
of  a negative correlation between
FUV - V color and Mg line strength originally proposed  by Burstein et al. (1988).
The present study provides the following possible explanation for the origin
of this ``Burstein relation''
(i.e., a correlation of Mg$_2$ with ($1550-V$) color).
More massive elliptical galaxies can retain more efficiently the
ejecta of supernovae so that they can finally have higher metallicities
of their stars (e.g., Arimoto \& Yoshii 1987). 
If more massive  elliptical galaxies
have shallower (i.e., more top-heavy) IMFs, then
they can have large $F_{\rm He}$ and thus show the stronger UV upturn. 
Therefore more massive (or luminous)  elliptical galaxies can have
higher metallicities and thus higher Mg$_2$ as well as the stronger UV upturn.
Given that the observed positive correlations between luminosities, Mg indices, and 
velocity dispersions in elliptical galaxies (e.g., Faber \& Jackon 1976; Bender et al. 1992),
more massive (or luminous)  elliptical galaxies can have
higher Mg$_2$, 
higher stellar velocity dispersion, and
the stronger UV upturn.
Thus,  the dependence of IMF slopes on galactic masses/luminosities can be
responsible for the origin of the Burstein relation.

Although Hoversten \& Glazebrook (2008) revealed that 
galaxies significantly fainter than the Galaxy show steeper IMFs,
it remain observationally unclear how IMF slopes depend on global galactic
properties and formation environments of galaxies.
Therefore it is too premature to conclude whether the CDS with  IMF variation 
is promising or not. As reviewed by Yi (2008), only a small fraction of
elliptical galaxies show the UV upturn, which needs an explanation.
If elliptical galaxies are formed from merging of smaller galaxies with different IMFs,
then the origin of the UV upturn would not be so simple as the CDS 
explains.

The observed strong correlation between Mg$_2$ and  $(1550-V$) color 
(Burstein et al. 1988)
suggests that metallicities play a role in the formation
of the correlation.
Figure 1 in Burstein et al. (1988)
also showed a  large difference in $(1550-V$) colors ($\sim 2.5$ mag) among
elliptical with different Mg$_2$  ranging from $\sim 0.20$ to $0.36$.
It is unclear whether this large difference can be {\it quantitatively}  explained 
by IMF variation alone, because the present study can not predict  $(1550-V)$
colors as a function of $F_{\rm He}$.
It would be possible that the combination
of high metallicities and large $F_{\rm He}$ can 
make $(1550-V)$ colors significantly bluer in elliptical galaxies. 
It is accordingly our future study to include chemical evolution and thereby to discuss
the dependences of $F_{\rm He}$ on metallicities in elliptical galaxies.

\subsection{Radial gradients of helium-rich stars in galactic spheroids}

Carter et al. (2011) have 
recently investigated radial gradients of the FUV excess
in 52 galaxies observed by $GALEX$ and found that some of them show a positive gradient in the
($FUV$ - $NUV$) color. They therefore suggested that the observed gradient can be
due to a helium abundance gradient and furthermore that the presence of the gradient
can place a strong limit on the importance of dry mergers 
in elliptical galaxy formation.
In order to discuss the origin of the observed radial gradients of 
($FUV$ - $NUV$) color, we have constructed a toy N-body model in which
a galactic spheroid is formed from merging of numerous SCs with
the mass fraction of HRSs being a free parameter. The details of the model
are given in the Appendix A.

Figure 7 shows time evolution of spatial distributions of HNSs and HRSs in a galactic
spheroids composed initially of numerous SCs for the standard model.
As SCs closely interact with one another in early evolution
phase of the galactic spheroid, smaller 
and less massive SCs are destroyed by the tidal field of the host spheroid
and by larger and more massive
SCs. The stars (HNSs) initially in less massive SCs are dispersed during
destruction of their host SCs and finally become field HNSs.  On the other hand,
massive SCs with $m_{\rm gmc} \ge M_{\rm thres}$ and thus with HRSs can not be destroyed
efficiently during early multiple SC interaction. These massive SCs can sink into
the inner region of the spheroid owing to dynamical friction against field stars (HNSs)
and then interact dynamically will other massive SCs there.
Consequently,  the massive SCs can get disintegrated there and 
their HRSs can be dispersed to become field HRSs there. Since the inner region
of the spheroid can be finally dominated by stars originally from HRSs of massive SCs,
the inner region can have a larger fraction of HRSs within $\sim 1 $ Gyr dynamical 
evolution.

Figure 8 shows how radial $F_{\rm He, t}$ gradients depend on models parameters,
$M_{\rm thres}$ and $\gamma$,  for a give $\beta$ (=1.7).
Clearly, all models shows negative gradients in the sense that inner regions 
of galactic spheroids have higher fractions of HRSs.
For example, $F_{\rm He, t}$ can be $\sim 0.08$ within the central 200 pc in the standard
model, though it is rather small ($\sim 0.01$) in the outer region ($R \sim1.8$ kpc).
The parameter $M_{\rm thres}$ can control the central value of $F_{\rm He, t}$ 
such that $F_{\rm He, t}$ can be larger for smaller $M_{\rm thres}$.
The $F_{\rm He, t}$ gradients can be smaller in models with $\gamma=0$ in which
there is no   mass-size relation for SCs. The derived rather flat radial distribution
is due to higher degrees of dynamical mixing of SCs with different mass in these
models.
Since $F_{\rm He}$ is proportional to $F_{\rm He, t}$ for a given set of IMFs,
these results on $F_{\rm He,t}$ gradients can be true for $F_{\rm He}$.
The present study accordingly
demonstrates that galactic spheroids can show negative
radial gradients of $F_{\rm He, t}$ and $F_{\rm He}$
and therefore  negative gradients of the strength of the UV upturn
(i.e., stronger in their inner regions).
The present study thus can clearly explain the above observational
result by Carter et al. (2011), 
if the  observed color gradients are  due largely to  helium
abundance gradients.

\section{Conclusions}

We have adopted a model in which (i) all stars are formed
as bound or unbound SCs from GMCs
and (ii) HRSs with $Y\ge 0.35$
can be formed from gas ejected from
AGB stars only in  massive SCs with $m_{\rm gmc} \ge M_{\rm thres}$.
Some SCs can become field stars in their host galaxies after disintegration
and others can become massive 
GCs or nuclear star clusters (or stellar nuclei) in galaxies.  
If massive SCs with HRSs are destroyed by their host galaxies,
then HRSs become helium-rich field stars in the galaxies.
We have investigated mass fractions of MS HRSs ($f_{\rm He}$ and $F_{\rm He}$) in 
GCs and galactic spheroids
and their dependences on IMFs, $M_{\rm thres}$, and GMCMFs
based on the  CDS (``cluster disintegration scenario'').
We have also investigated the radial gradients of $F_{\rm He}$
in galaxies with mass-size relations of GMCs
by using N-body simulations on dynamical evolution of multiple cluster systems.
We summarize our principle results as
follows. \\

(1) The mass ratios  ($f_{\rm m, agb}$) of gaseous ejecta
from  massive AGB stars with $3{\rm M}_{\odot} \le m_{\rm I} \le 8 {\rm M}_{\odot}$
to initial total masses of SCs are typically 
$\sim 0.08$
for a canonical IMF (${\alpha}_1 = 2.35$) with the
reasonable lower and upper cut-off masses.
If these gas can be efficiently converted into new stars 
with enhanced helium abundances (i.e., HRSs) in SCs,
then the mass fractions of MS HRSs  among all MS stars ($f_{\rm He}$)
are $\sim 0.08$  for canonical IMFs
and a SC age of 12 Gyr.
Only if original SCs have top-heavy IMFs (e.g., ${\alpha}_1 \sim 1.5$),
then $f_{\rm He}$ can be larger than  0.3
for $\alpha_2 \ge  1.85$. \\

(2) The initial number and mass fractions of massive GCs (MGCs) with  HRSs
($F_{\rm n, mgc}$ and $F_{\rm m, ngc}$, respectively)
among the Galactic GCs formed from GMCs with $m_{\rm gmc} \ge 10^6 {\rm M}_{\odot}$
is $\sim 0.16$ and $0.66$, respectively,  for $M_{\rm th}=10^7 {\rm M}_{\odot}$
and $\beta =1.7$. Both $F_{\rm n, mgc}$ and $F_{\rm m, mgc}$
are lager for smaller $M_{\rm th}$ and smaller $\beta$.
MGCs with $m_{\rm gc}= 5 \times 10^6 {\rm M}_{\odot}$ 
can sink into the Galactic center owing to dynamical friction
to disappear from the halo region within
13 Gyr,
if the initial radii from the Galactic center are less than 3.4 kpc.
About 37\% (49\%)  of the Galactic GCs 
with $m_{\rm gc} \ge 5 \times 10^6 {\rm M}_{\odot}$ 
($m_{\rm gc} \ge  10^7 {\rm M}_{\odot}$)
have already been lost from the halo owing to efficient dynamical friction
of these massive GCs. Thus, the Galactic halo
initially could have a larger fraction of  MGCs. \\

(3) The mass fractions of HRSs among all stars 
($F_{\rm He, t}$) in galactic spheroids are 
$\sim 0.05$ for $M_{\rm thres}=10^7 {\rm M}_{\odot}$,   $\beta =1.7$, 
$f_{\rm m, agb}=0.1$, and ${\epsilon}_{\rm bsc}=1$.  $F_{\rm He, t}$ of galactic spheroids
are  larger for smaller $M_{\rm th}$ and
smaller $\beta$ for a given IMF.
The mass fractions of MS HRSs among all MS stars
($F_{\rm He}$) in galactic spheroids are
 0.06 for canonical IMFs ($\alpha_1=\alpha_2=2.35$),
$M_{\rm th}=10^7 {\rm M}_{\odot}$, $\beta=1.7$, and a galaxy age of $\sim 12$ Gyr.
$F_{\rm He}$ in galactic spheroids
can be  larger for smaller $M_{\rm th}$ and
smaller $\beta$ for a given IMF,  and it can be significant ($>0.1$),
if original SCs (i.e., building blocks of the spheroids)
have top-heavy IMF (e.g. $\alpha_1 <  2$).
These results suggest that if the observed UV upturn 
in bright elliptical galaxies  is due to the larger fractions 
($\sim 10$\%)  
of HRSs  within them, 
the IMFs need to be top-heavy.
If ${\epsilon}_{\rm bsc}$ is as small as $\sim 0.1$, then
bright elliptical galaxies are unlikely to show the UV upturn owing to the 
small fractions of HRSs. \\

(4) The Galactic bulge can have a larger $F_{\rm He}$
if it had a top-heavy IMF at its formation in the CDS. 
Given a number of suggestions on the top-heavy IMF of the bulge
by chemical evolution models of the bulge,
the observed possible presence of helium-rich stars 
with $Y \sim 0.35$ in the bulge (e.g., Nataf et al. 2011)
can be understood in the context of  the CDS.
HRSs in  MGCs that
were initially located in the Galactic halo and had  sunk into
the bulge can be currently observed as field HRSs  in the bulge,
though the contribution of such stars to the entire helium-rich population
is rather minor. It is doubtlessly worthwhile for observational studies
to investigate whether or not the bulge has a larger fraction of HRSs
in the inner regions. \\

(5) If $M_{\rm th}$ and $\beta$ are constant in 
elliptical galaxies with different physical
properties,  
then elliptical galaxies with shallower (or more top-heavy) IMFs can show
larger fraction of HRSs.
Therefore, if more massive/luminous elliptical galaxies 
have shallower IMFs, then they can have larger fraction of
HRSs.
More massive elliptical galaxies can retain more efficiently the
ejecta of supernovae so that they can finally have higher metallicities
of their stars (e.g., Arimoto \& Yoshii 1987). 
Therefore,
if the Burstein relation (a correlation between UV flux and Mg index;
Burstein et al. 1988) in bright elliptical galaxies is due to the  
correlation between $F_{\rm He}$  and Mg indices,
then the relation  can be understood in terms of shallower IMF slopes ($\alpha_1$) in
more massive/luminous elliptical galaxies. \\

(6) Galactic spheroids  can show negative radial gradients
of $F_{\rm He,t}$ and $F_{\rm He}$  (i.e., higher in inner parts)
for a reasonable set of model parameters. This is mainly because
massive SCs with HRSs 
can sink rapidly into the inner regions of galaxies owing to dynamical friction
and disintegrate there. 
The HRSs initially in the  SCs can be dispersed to become field HRSs there after
disintegration of the host SCs.
Therefore they can show the stronger UV upturn
in their inner regions, if $F_{\rm He}$ can determine
the strengths of the UV upturn. 
The observed spatial distribution of the FUV excess
in elliptical galaxies can be understood in the context of formation and evolution
of MGCs with HRSs. 
The central regions of galactic spheroids can be composed of 
two different stellar populations
with normal helium abundances and enhanced ones (i.e., HNSs and HRSs).\\


Thus, an advantage of the CDS is that both  (i) the absence or presence
of the UV upturn in galactic spheroids  and (ii) correlations between 
the strength of the UV upturn  and galactic
properties (e.g., Mg$_2$ index) can be discussed in the context of IMF slopes
of galaxies in a self-consistent manner.  A disadvantage  of the CDS is that
it is unclear whether and in what physical conditions
${\epsilon}_{\rm bsc}$ can be $\sim 1$ for massive GMCs
in galaxies.
The formation of HRSs is possible,
if helium-rich gas from AGB stars is converted into new
stars without dilution of the gas by  ISM with normal helium abundances.
In the CDS,  the inner regions of massive SCs 
are assumed to be the production sites of HRSs.
It would be possible that AGB ejecta
can be converted into new cold gas clouds outside SCs and consequently into new
stars in deep potential wells of galactic spheroids, if there is no/little
cold gas with normal helium abundances there.
We need to investigate where and how AGB ejecta from field stars and SCs can be converted 
into new stars in galaxies using a more sophisticated chemodynamical  simulation
in our future studies.

\section{Acknowledgment}
I am  grateful to the  anonymous referee for valuable comments
which contribute to improve the present paper.
KB acknowledge the financial support of the Australian Research Council
throughout the course of this work.

\appendix

\section{N-body models for multiple SC evolution}

We investigate dynamical evolution of a galactic spheroid composed of
numerous SCs using collisionless N-body simulations
based on an idealized  model for galactic spheroids.
The main aim of this numerical investigation is
to illustrate the radial gradients of the mass fractions
of HRSs ($F_{\rm He, t}(R)$) in the spheroids.
This idealized model is used only in the present preliminary study for the origin
of helium-rich stars in galactic spheroids, and more sophisticated 
models including chemical evolution due to gaseous ejection
from supernovae and AGB stars will be used in our future studies.
However, we consider that
this ``toy''  model enables us to grasp some essential ingredients of
the formation and evolution of radial $F_{\rm He, t}$ gradients in galactic spheroids.

A galactic spheroid has an initial
total  mass $M_{\rm gal}$ and a size $R_{\rm gal}$ 
and composed of SCs with the total number of $N_{\rm sc, t}$.
The radial distribution of SCs are described by
a Plummber  profile
(e.g., Binney \& Tremaine 1987) with the scale-length of $0.2 R_{\rm gal}$.
Each SC 
with a  mass $m_{\rm sc}$ and a  size $r_{\rm sc}$ is  composed of many
stars and has a Plummber density profile
with the scale-length of $0.2 r_{\rm sc}$.
A galactic spheroid is assumed to be initially in dynamical equilibrium,
so the  velocities of each SCs
is given according to the velocity dispersion profile of the adopted Plummer model.

A SC is composed of HNSs and HRSs and these two populations have different
initial distributions within the SC.
Recent numerical simulations have shown that secondary star formation
from AGB ejecta can proceed mostly in the central regions of MGCs 
(e.g., D'Ercole et al. 2008; Bekki 2010, 2011). Guided by these results,
HNSs and HRSs are located at  $0.1r_{\rm sc} \le R \le r_{\rm sc}$ 
and $R < 0.1 r_{\rm sc}$, respectively. The mass fraction of HRSs are 
calculated for a given $\beta$ and $M_{\rm thres}$. The total particle
number in a SC is proportional to $m_{\rm sc}$ so that masses of stellar particles
can be all the same.

A progenitor GMC for a bound/unbound SC
has a mass $m_{\rm gmc}$ and a size $r_{\rm gmc}$.
If we use the observed relation between
mass densities and sizes of GMCs discovered
by Larson (1981)
and the observed typical mass and size of GMCs in
the Galaxy (e.g., Solomon et al. 1979),
then $m_{\rm gmc}-r_{\rm gmc}$ relation can be described as follows:
\begin{equation}
r_{\rm gmc}
=40 \times  (\frac{m_{\rm gmc}}{5 \times 10^5  {\rm M}_{\odot} })^{\gamma}
{\rm pc},
\end{equation}
where $\gamma \sim 0.5$.
We investigate models with different $\gamma$, because initial SCs can have 
mass-size relations different from those of GMCs. 
We consider that when SCs are formed,  SCs have a  mass-size
relation similar to the above one and thus assume that $r_{\rm sc}$ can be
the same as $r_{\rm gmc}$ described above for a given $m_{\rm gmc}$.

We mainly present the results of the ``standard'' dynamical models in which 
$M_{\rm gal}=10^{9} {\rm M}_{\odot}$,  $R_{\rm gal}=2$ kpc,
$\beta=1.7$, 
$M_{\rm thres}=10^7 {\rm M}_{\odot}$
$m_{\rm gmc, l}=10^4 {\rm M}_{\odot}$,
$m_{\rm gmc, u}=10^8 {\rm M}_{\odot}$,
and $\gamma=0.5$, though we investigate models with different parameters.
In order to perform numerical simulations,
we use the latest version of GRAPE
(GRavity PipE, GRAPE-DR), which is the special-purpose
computer for gravitational dynamics (Sugimoto et al. 1990).
The total particle number used in a simulation is $\sim 200000$,
which we consider to be reasonable for this preliminary investigation.
The gravitational softening length is set to be 20 pc for all models.
We focus only on the final radial gradients of $F_{\rm He, t}$ in galactic spheroids
formed from dynamical evolution of numerous SCs.

\newpage

\begin{deluxetable}{cc}
\footnotesize
\tablecaption{Meaning of acronyms
\label{tbl-1}}
\tablewidth{0pt}
\tablehead{
\colhead{  Acronyms } &
\colhead{  Meanings  } }
\startdata
CDS  & Cluster disintegration scenario  \\
SC  & Star cluster  \\
GC  & Globular cluster  \\
HNS  & Helium-normal stars   \\
HRS  & Helium-rich stars   \\
MS  & Main-sequence (stars)   \\
MGC  & Massive GCs (i.e., SCs with HRSs)   \\
GMCMF  & GMC mass function   \\
\enddata
\end{deluxetable}

\begin{deluxetable}{cc}
\footnotesize
\tablecaption{Model parameters and their physical meanings 
\label{tbl-2}}
\tablewidth{0pt}
\tablehead{
\colhead{  Parameters } &
\colhead{  Physical meanings  } }
\startdata
$\alpha_1$  & The IMF slope for HNSs   \\
$\alpha_2$  & The IMF slope for HRSs   \\
$m_l,1$  & The lower mass cut-off of the IMF for HNSs  \\
$m_u,1$  & The upper  mass cut-off of the IMF for HNSs \\
$m_l,2$  & The lower mass cut-off of the IMF for HRSs \\
$m_u,2$  & The upper  mass cut-off of the IMF for HRSs \\
$\beta$  & The slope of the GMC mass function (GMCMF) \\
$M_{\rm thres}$  & The threshold mass for MGC formation \\
$\epsilon_{\rm bsc}$  &  The formation efficiency of bound SCs within GMCs 
(fixed)\\
\enddata
\end{deluxetable}
\vspace{10cm}

\newpage

\vspace{10cm}
\begin{deluxetable}{cc}
\footnotesize
\tablecaption{The definition of physical quantities  
investigated in the present study
\label{tbl-3}}
\tablewidth{0pt}
\tablehead{
\colhead{  Quantity } &
\colhead{  Meaning  } }
\startdata
$f_{\rm m, He}$  & The mass fraction of fresh helium gas in a SC \\ 
$f_{\rm m, agb}$  & The mass fraction of AGB ejecta in a SC \\ 
$f_{\rm He}$  & The mass fraction of MS HRSs in a SC \\ 
$F_{\rm n, mgc}$  & The number fraction of MGCs with HRSs in all genuine GCs\\ 
$F_{\rm m, mgc}$  & The mass fraction of MGCs with  HRSs in all genuine GCs \\ 
$F_{\rm t, He}$  & The mass fraction of HRSs in a spheroid \\ 
$F_{\rm He}$  & The mass fraction of MS HRSs in a spheroid \\ 
\enddata
\end{deluxetable}
\vspace{10.0cm}

\newpage

\begin{figure}
\epsscale{1.0}
\plotone{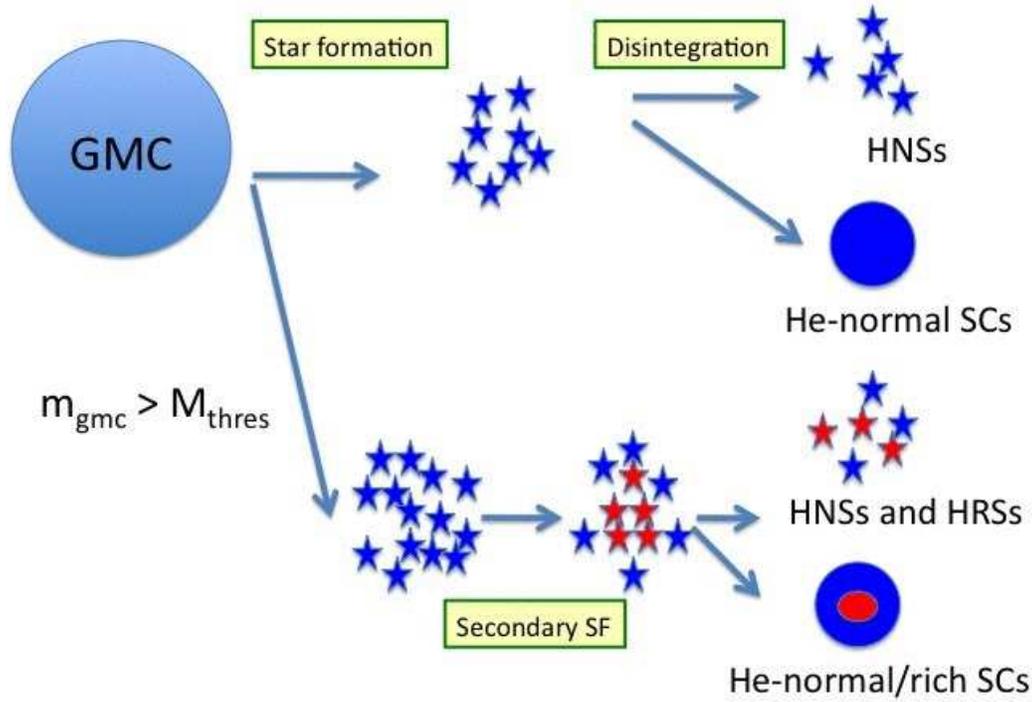}
\figcaption{
An  illustration for  the whole picture of the star formation processes
from GMCs in the CDS. The evolution of newly formed
HNSs (blue stars) depends on the masses of their
host GMCs ($m_{\rm gmc}$).  HNSs formed in lower mass GMCs
with $m_{\rm gmc} \le M_{\rm thres}$ can not form HRSs (red stars) from gaseous
ejecta of their AGB stars owing to the shallow gravitational potentials
of the SCs.
The HNSs can finally evolve either field stars after disintegration
of their host SCs or into bound SCs with only HNSs. On the other hands,
if $m_{\rm gmc} > M_{\rm thres}$,
then  gaseous ejecta of AGB stars among HNSs can be converted into
new HRSs (red stars) owing to deeper gravitational potentials of the SCs.
The stars in the SCs can become either field HNSs/HRSs or main components
of massive SCs with HRSs depending on physical properties of the SCs
(i.e., on whether the SCs become disintegrated).
The field HRSs in galactic spheroids are responsible for the UV upturn
in the CDS.
\label{fig-1}}
\end{figure}

\begin{figure}
\epsscale{0.8}
\plotone{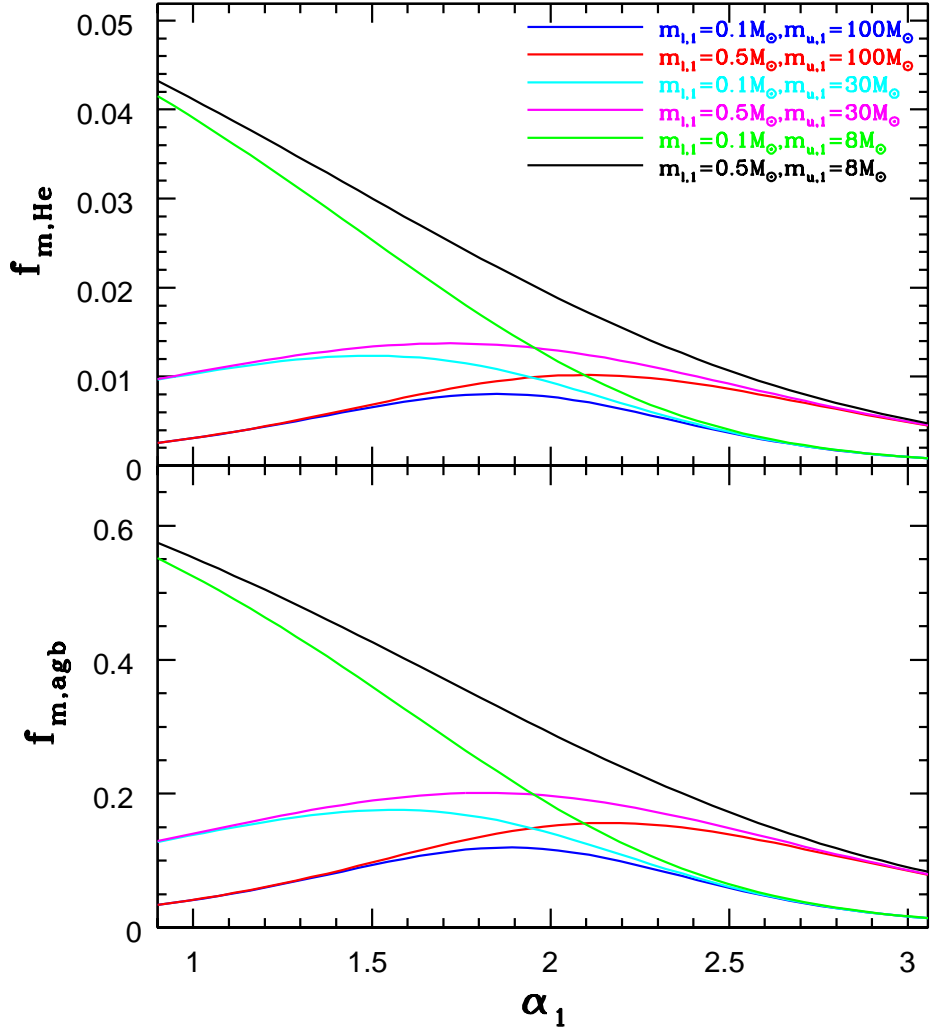}
\figcaption{
The dependences of the mass fractions of fresh helium gas ($f_{\rm m, He}$; upper)
and those of the AGB ejecta  ($f_{\rm m, agb}$; lower) in SCs on the IMF slopes
of SCs ($\alpha_1$) for different lower ($m_{\rm l,1}$) and upper ($m_{\rm u,1}$)
mass cut-offs:
$m_{\rm l,1}=0.1 {\rm M}_{\odot}$ and $m_{\rm u,1}=100 {\rm M}_{\odot}$ (blue),
$m_{\rm l,1}=0.5 {\rm M}_{\odot}$ and $m_{\rm u,1}=100 {\rm M}_{\odot}$ (red),
$m_{\rm l,1}=0.1 {\rm M}_{\odot}$ and $m_{\rm u,1}=30 {\rm M}_{\odot}$ (cyan),
$m_{\rm l,1}=0.5 {\rm M}_{\odot}$ and $m_{\rm u,1}=30 {\rm M}_{\odot}$ (magenta),
$m_{\rm l,1}=0.1 {\rm M}_{\odot}$ and $m_{\rm u,1}=8 {\rm M}_{\odot}$ (green),
and $m_{\rm l,1}=0.5 {\rm M}_{\odot}$ and $m_{\rm u,1}=8 {\rm M}_{\odot}$ (black).
Here $\alpha_1=2.35$ corresponds to the Salpeter IMF.
\label{fig-2}}
\end{figure}

\begin{figure}
\epsscale{0.8}
\plotone{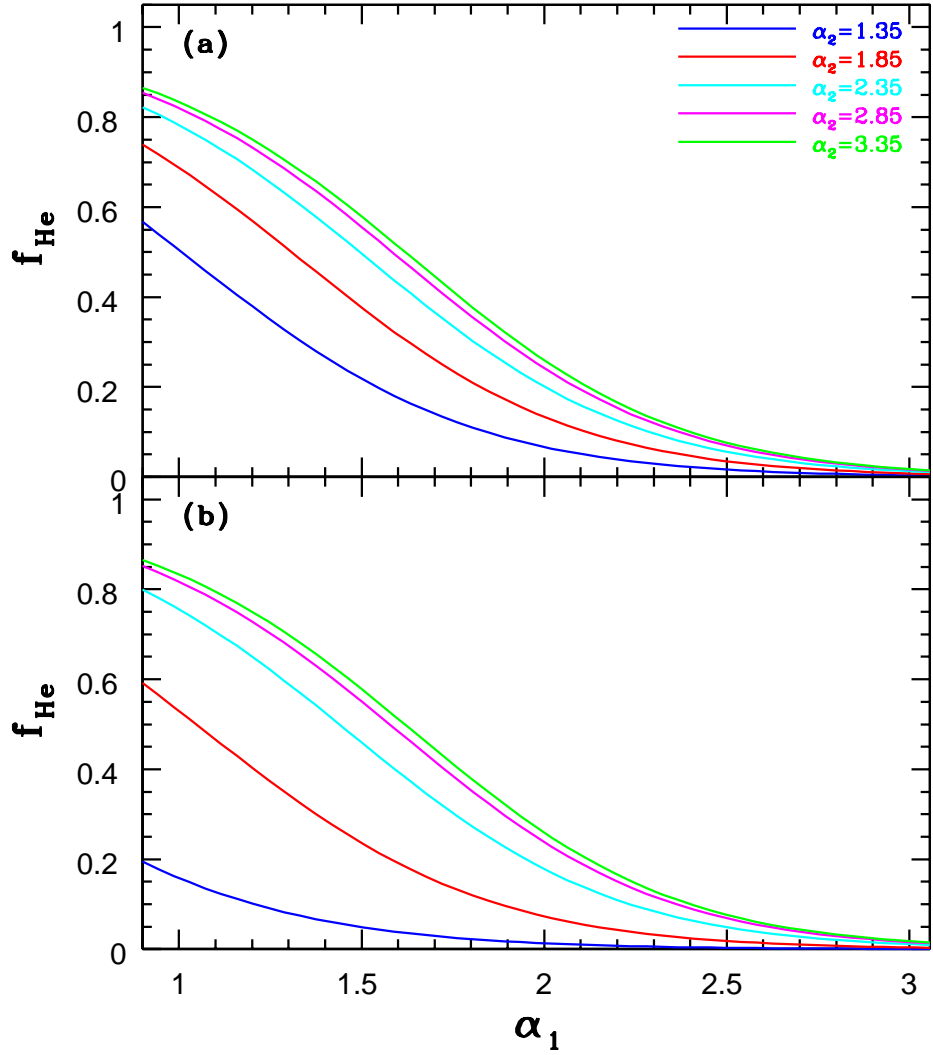}
\figcaption{
The dependences of mass fractions of MS HRSs ($f_{\rm He}$) on $\alpha_1$ in SCs for
$m_{\rm l,2}=0.1 {\rm M}_{\odot}$ and $m_{\rm u,2}=8 {\rm M}_{\odot}$ (upper),
and $m_{\rm l,2}=0.1 {\rm M}_{\odot}$ and $m_{\rm u,2}=100 {\rm M}_{\odot}$ (lower)
for different IMF slopes for HRSs ($\alpha_2$):
$\alpha_2=1.35$ (blue),
$\alpha_2=1.85$ (red),
$\alpha_2=2.35$ (cyan),
$\alpha_2=2.85$ (magenta),
and $\alpha_2=3.35$ (green).
Here $\alpha_1=2.35$ corresponds to the Salpeter IMF.
\label{fig-3}}
\end{figure}

\begin{figure}
\epsscale{0.8}
\plotone{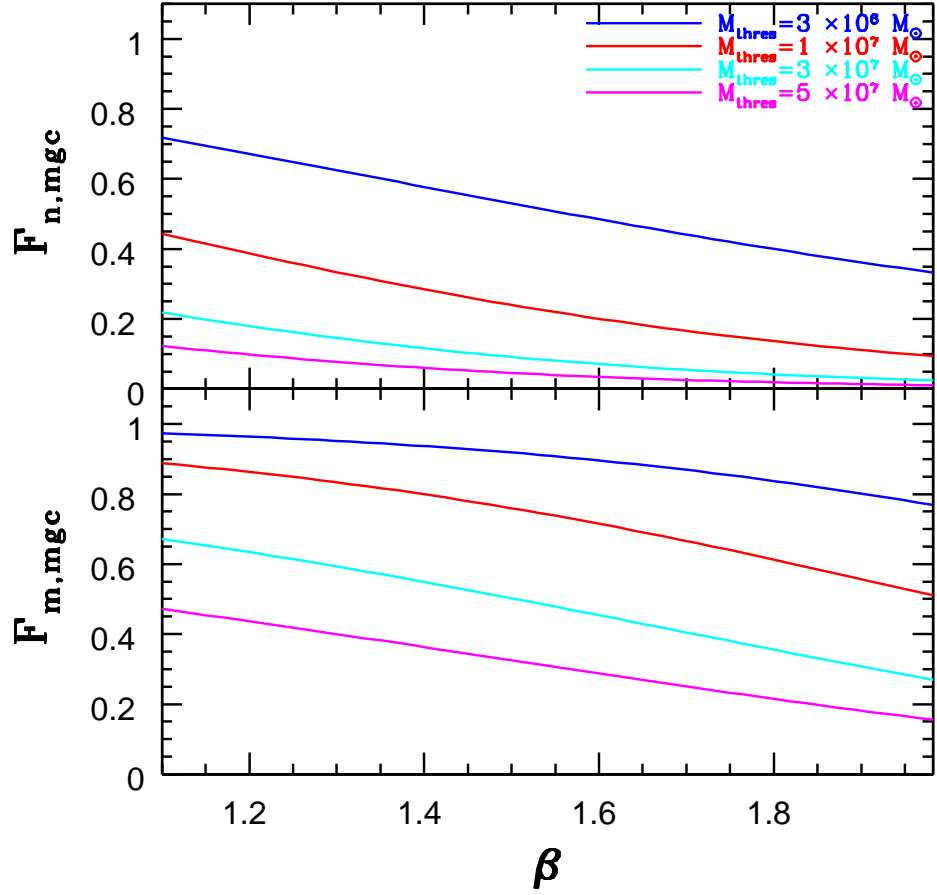}
\figcaption{
The dependences of number and mass fractions of massive GCs
with HRSs ($F_{\rm n, mgc}$ and $F_{\rm m, mgc}$, respectively)
in the Galactic ``genuine'' GCs 
(formed from GMCs with 
$m_{\rm gmc} \ge 10^6  {\rm M}_{\odot}$) on  GMCMF slopes ($\beta$)
for different threshold masses for MGC formation ($M_{\rm thres}$):
$M_{\rm thres}=3 \times 10^6 {\rm M}_{\odot}$ (blue),
$M_{\rm thres}=1 \times 10^7 {\rm M}_{\odot}$ (red),
$M_{\rm thres}=3 \times 10^7 {\rm M}_{\odot}$ (cyan),
and $M_{\rm thres}=5 \times 10^7 {\rm M}_{\odot}$ (magenta).
\label{fig-4}}
\end{figure}

\begin{figure}
\epsscale{0.8}
\plotone{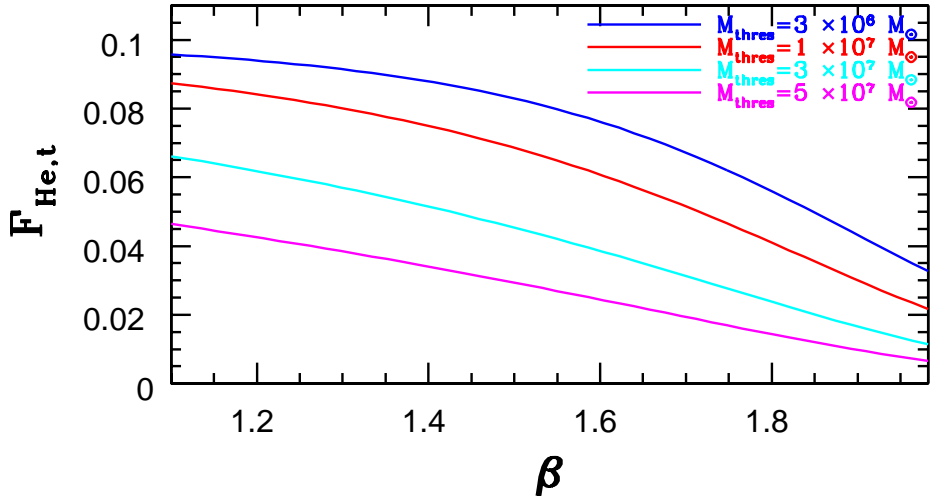}
\figcaption{
The dependences of 
the  mass fractions of HRSs ($F_{\rm He, t}$) in galactic spheroids
on  GMCMF slopes ($\beta$)
for canonical IMFs ($\alpha_1=2.35$) and
different threshold masses for MGC formation ($M_{\rm thres}$):
$M_{\rm thres}=3 \times 10^6 {\rm M}_{\odot}$ (blue),
$M_{\rm thres}=1 \times 10^7 {\rm M}_{\odot}$ (red),
$M_{\rm thres}=3 \times 10^7 {\rm M}_{\odot}$ (cyan),
and $M_{\rm thres}=5 \times 10^7 {\rm M}_{\odot}$ (magenta).
\label{fig-5}}
\end{figure}

\begin{figure}
\epsscale{0.7}
\plotone{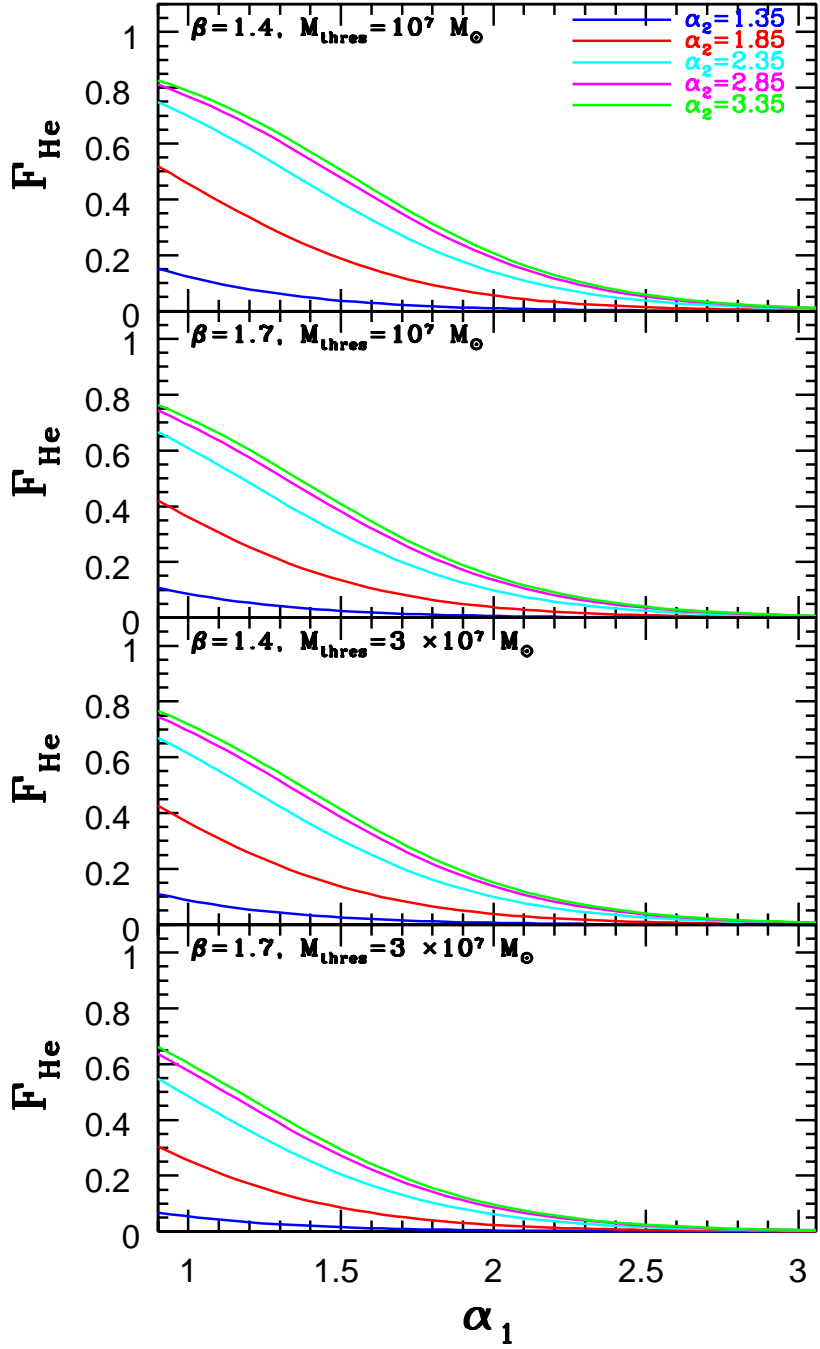}
\figcaption{
The dependences of  mass fractions of MS HRSs in galactic spheroids
($F_{\rm He}$) on $\alpha_1$
in the models with
$\beta=1.4$ and $M_{\rm thres}=10^7  {\rm M}_{\odot}$ (top),
$\beta=1.7$ and $M_{\rm thres}=10^7  {\rm M}_{\odot}$ (the second from top),
$\beta=1.4$ and $M_{\rm thres}=3 \times 10^7  {\rm M}_{\odot}$ (the second from bottom),
$\beta=1.7$ and $M_{\rm thres}=3 \times 10^7  {\rm M}_{\odot}$ (bottom)
for different $\alpha_2$:
$\alpha_2=1.35$ (blue),
$\alpha_2=1.85$ (red),
$\alpha_2=2.35$ (cyan),
$\alpha_2=2.85$ (magenta),
and  $\alpha_2=3.35$ (green).
\label{fig-6}}
\end{figure}

\epsscale{1.0}
\begin{figure}
\plotone{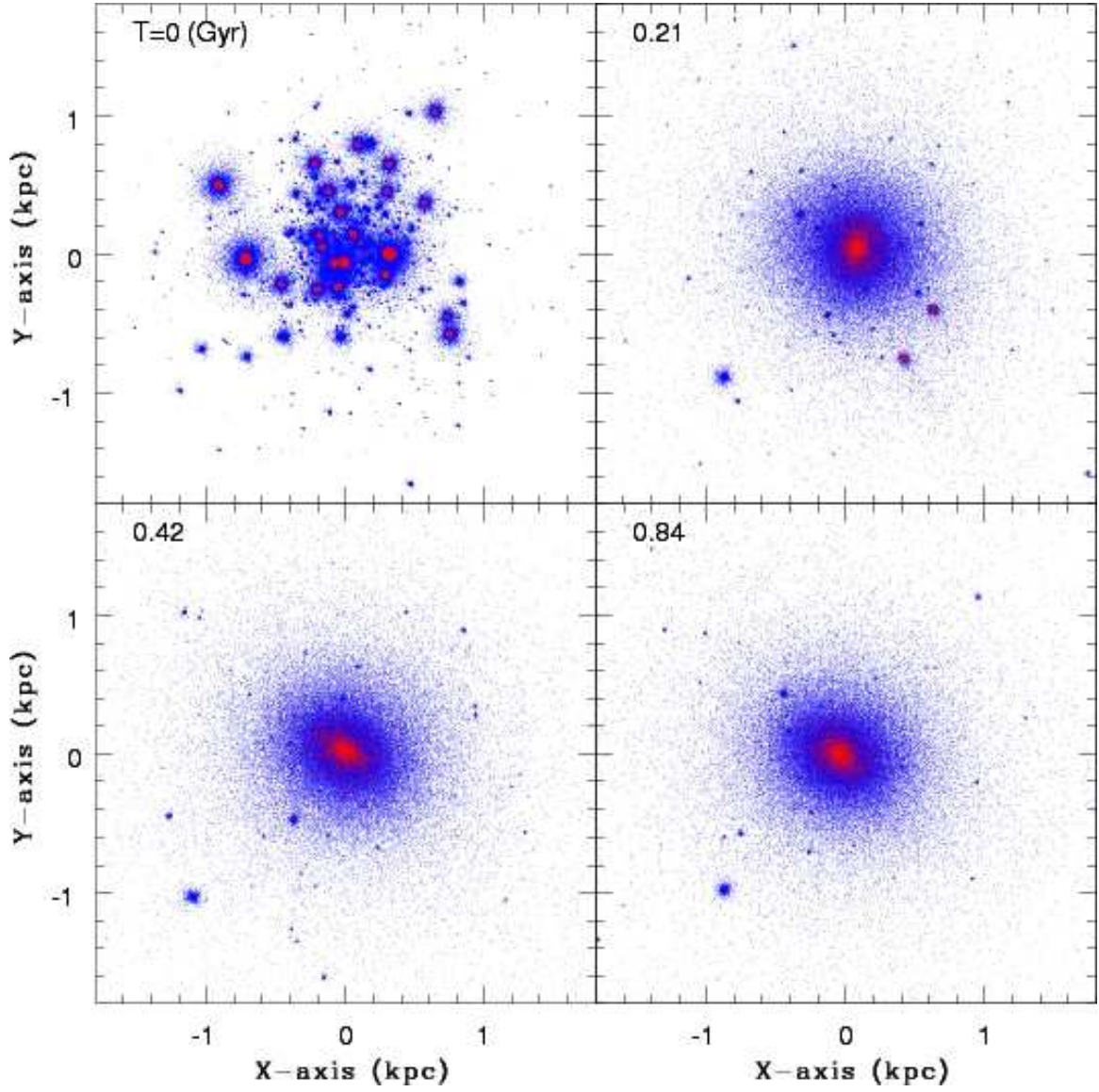}
\figcaption{
Time evolution of the distributions of HNSs (blue) and HRSs (red) projected onto
the $x$-$y$ plane for the standard model. The time ($T$
in units of Gyr)  that has elapsed since
the simulation starts is shown in the upper left corner of each panel.
\label{fig-7}}
\end{figure}

\begin{figure}
\plotone{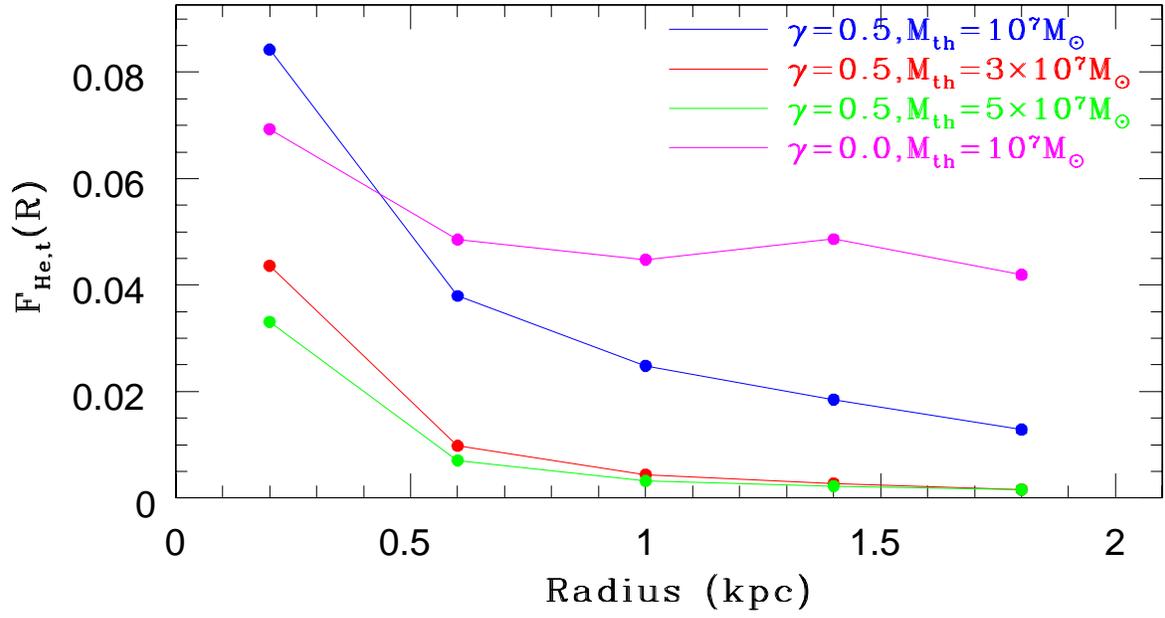}
\figcaption{
The radial gradients of mass fractions of HRSs ($F_{\rm He, t}$) in models
with different parameters:
$\gamma=0.5$ and $M_{\rm thres}=10^7 {\rm M}_{\odot}$ (blue),
$\gamma=0.5$ and $M_{\rm thres}=3 \times 10^7 {\rm M}_{\odot}$ (red),
$\gamma=0.5$ and $M_{\rm thres}=5 \times 10^7 {\rm M}_{\odot}$ (green),
and $\gamma=0.0$  and $M_{\rm thres}=10^7 {\rm M}_{\odot}$ (magenta).
\label{fig-8}}
\end{figure}

\end{document}